\documentclass[fleqn,usenatbib]{mnras}

\usepackage{newtxtext,newtxmath}

\usepackage[T1]{fontenc}

\DeclareRobustCommand{\VAN}[3]{#2}
\let\VANthebibliography\thebibliography
\def\thebibliography{\DeclareRobustCommand{\VAN}[3]{##3}\VANthebibliography}

\usepackage{graphicx}	
\usepackage{amsmath}	
\usepackage{threeparttable}
\usepackage[dvipsnames]{xcolor}
\usepackage{float}
\usepackage{comment}

\definecolor{mygreen}{RGB}{0, 100, 20}

\title[Simulations of Restarted Jets - I. Dynamics]{Numerical Simulations of Restarted Jets - I. Dynamics}

\author[Prathamesh Ratnaparkhi et al.]{
Prathamesh Ratnaparkhi$^{1}$\thanks{E-mail: prathamesh.ratnaparkhi@iucaa.in},
Dipanjan Mukherjee$^{1}$,
Gianluigi Bodo$^{2}$,
 Paola Rossi$^{2}$,
 Marisa Brienza$^{3}$\newauthor
 and Raffaella Morganti$^{4,5}$
\\
$^{1}$Inter-University Centre for Astronomy and Astrophysics, Pune-411007, India\\
$^{2}$INAF, Osservatorio Astrofisico di Torino, Strada Osservatorio 20, I-10025 Pino Torinese, Italy\\
$^{3}$Istituto Nazionale di Astrofisica (INAF) - Istituto di Radioastronomia (IRA), via Gobetti 101, 40129, Bologna, Italy\\
$^4$ASTRON, the Netherlands Institute for Radio Astronomy, Oude Hoogeveensedijk 4, 7991 PD, Dwingeloo, The Netherlands\\
$^5$Kapteyn Astronomical Institute, University of Groningen, Postbus 800, 9700 AV Groningen, The Netherlands
}

\date{Accepted XXX. Received YYY; in original form ZZZ}

\pubyear{\the\year{}}

\begin{document}
\label{firstpage}
\pagerange{\pageref{firstpage}--\pageref{lastpage}}
\maketitle

\begin{abstract}
We performed high-resolution three-dimensional relativistic magnetohydrodynamic (RMHD) simulations of restarted jets evolving within the cavity of a previous jet episode, using the {\small PLUTO} code. The simulations cover a spatial domain of $50\;\mbox{kpc}$ with a resolution of $40\;\mbox{pc}$. Three suites of simulations were performed to understand the impact of jet power, magnetic fields and quiescence time on the evolution of restarted jets. The restarted jets undergo a complex, multi-stage evolution, with the remnant cocoon relaxing from an initially conical structure to a mushroom-shaped morphology via an intermediate cylindrical phase. As the cocoon of the initial jet expands, dense material entrained by fluid instabilities such as Kelvin-Helmholtz and Rayleigh-Taylor significantly alter the conditions within it. As a result, the interaction of the restarted jet with the cocoon is markedly different from that of the initial jet with the ambient medium. In particular, we find that the restarted jet propagates ballistically through the rarefied remnant cocoon, without creating prominent backflows. Deceleration of the jet and associated backflows are observed only when it encounters dense structures. The structure and strength of the shocks in restarted jets are affected by jet power, magnetic field strength, and quiescence time. Finally, we discuss the implications of the dynamics for observed properties of radio galaxies.
\end{abstract}




\section{Introduction}

Observations of jetted radio galaxies provide strong evidence of intermittent jet activity, with clear signatures of a complete turn-off and a restart of the jet. A major step in this direction was the systematic identification of double–double radio galaxies (DDRGs)  \cite{2000MNRAS.315..371S,2019A&A...622A..13M}, with a younger pair of inner radio lobes propagating within an older, more extended pair of outer lobes. Besides traditional DDRGs, subsequent studies have revealed a broader phenomenology associated with restarted activity. Several sources previously classified as single-episode radio galaxies have been shown to host faint, diffuse large-scale emission, plausibly linked to earlier phases of jet activity \citep[e.g.,][]{2013ApJ...765L..11S}. In other systems, the presence of compact inner jets or lobes suggests very recent reactivation of the active galactic nuclei (AGN) \citep{2012ApJ...760...77A}. More complex and less symmetric morphologies have also been identified, including misaligned jet axes \citep[e.g.,][]{2013MNRAS.436..690S, 2017A&A...603A.131H}, evidence for more than two distinct episodes of activity \citep[e.g.,][]{2007MNRAS.382.1019B, 2023MNRAS.525L..87C}, and highly irregular or disturbed structures \citep[e.g.,][]{2024A&A...691A.287N}. Besides, morphological signatures of restarted activity, spatial variation of spectral index in radio lobes \citep{2020A&A...638A..29B}, core prominence and other criteria \citep{2020A&A...638A..34J, 2024A&A...691A.287N} also aid identification of episodic activity. 

Dynamical modelling and spectral diagnostics provide the primary means of estimating the durations of active and quiescent phases in restarted jet systems. They reveal an extraordinary diversity of time-scales, from a  few Myrs to hundreds of Myrs. At the shortest end, very young restarted sources have been identified on parsec scales with very short restart times, e.g., $\sim 10$ yr for 3C~84, $82 \pm 17\;\mbox{yr}$ for TXS~0128+554, and several others, as reported in \cite{Nyland20a}. At intermediate scales, systems such as 3C~293 exhibit two to three distinct jet episodes ($\sim 0.17 - 0.27$ Myr), with the inner lobes ($\sim 2\;\mbox{kpc}$) embedded within $\sim 100\;\mbox{kpc}$ outer lobes \citep{2022A&A...658A...6K}. Similarly, \cite{2020A&A...634A...9M} reports episodes of duration $12\;\mbox{Myr}$ and $3\;\mbox{Myr}$, separated by quiescence time of $9\;\mbox{Myr}$ for Fornax A. At the opposite extreme, some double–double radio galaxies show outer lobes with spectral ages $\gtrsim 200\;\mbox{Myr}$ and inner doubles younger than $\sim 30\;\mbox{Myr}$ \citep{2007MNRAS.378..581J}. Broader samples confirm that restarted sources populate a wide continuum of evolutionary stages, spanning several orders of magnitude in size, power, and characteristic time-scales \citep{2024Galax..12...11M, 2024A&A...691A.287N}. Despite the wide range of timescales, the observational detection of restarted jets is biased against very short interruption timescales as the lobes of the new jet might merge quickly with the old lobes \citep{Elley.M.2026.flickering_jets,2026MNRAS.546ag131E,2019A&A...622A..13M}.

 Population studies indicate that once the jets are switched off, radio emission fades rapidly, requiring a substantial population of short-lived or intermittently active sources to account for the observed numbers of remnant and restarted galaxies \citep{2020MNRAS.496.1706S}. Indeed, a study of compact steep spectrum sources by \cite{2023MNRAS.522.3877O} has found evidence for restarted activity, suggesting a power law distribution of source ages (short duty cycles). At the same time, statistical analyses show that the host galaxies of restarted radio sources are largely indistinguishable from those of the general radio-galaxy population, implying that jet interruption and reactivation are not exceptional events but rather a common phase in the life cycle of radio-loud AGN \citep{2019A&A...622A..13M, 2024A&A...691A.287N}. A population study by \cite{2021Galax...9..122J} also finds that the restarted activity for galaxies of the Lockman hole might be dominated by secular processes like stability of the accretion disk rather than fuelling of the central black hole. Taken together, these results suggest that episodic jet activity spans a broad and continuous range of physical scales and time-scales. Recurrent jet launching should be viewed as a normal aspect of galaxy evolution, with important consequences for AGN feedback. For example, observational studies \citep{2010ApJ...715..172T, 2022MNRAS.509.4997Z} have indeed reported correlations between star formation activity and jet episodes of the host galaxy.

Motivated by these observational findings, several numerical studies have investigated the dynamics of restarted jets. The very first simulation of restarted jets by \cite{1991ApJ...369..308C} found that the relaunched jet was over-dense compared to the cocoon material left by the previous jet. The advance speed of the new jet was much faster than that of the first jet. Similar conclusions have also been drawn by \cite{2014MNRAS.439.3969W}, which also considered the effect of different jet compositions on the restarted activity using magnetohydrodynamic (MHD) simulations in 2.5 dimensions. A follow-up study \citep[][]{2020MNRAS.497.3638W} examined the visibility criteria for double-double sources with various jet models, magnetic field morphologies, and viewing angles. The authors found that for a double-double source to be visible, the next jet must be launched within $~4$\% of the lifetime of the earlier jet. Another study by \citet[][]{2012ApJ...750..166M} found a strong impact of the cluster weather on the visible morphology of a restarted jet. Similarly, hydrodynamic simulations by \cite{2018MNRAS.480.5286Y} demonstrated that both the ambient environment and jet intermittency strongly influence jet collimation, lobe morphology, and the formation of multiple hotspots. RMHD simulations by \citet{2017MNRAS.469.4957B} have explored the dominant mechanisms of magnetic energy dissipation in single- and multi-episode jets through 2D and 3D simulations. Besides, restarted jets, a few studies \citep{2001ApJ...554..261C,2002MNRAS.332..271R,chen23a,2025PASA...42..152S} have explored the dynamics of the remnant cocoon formed after the jet is turned off through numerical simulations.

Future and ongoing radio missions, such as SKA, EMU, and MeerKAT, are expected to discover a large number of remnant and restarted radio galaxies. Thus, a robust, numerically based model is required for accurate interpretation of the observations. Due to the large length and time-scales, as well as the separation of scales between relativistic particles and bulk flows, there has not been commensurate progress in numerical modelling compared to observations. 

Existing numerical studies have explored the dynamics of restarted and remnant radio sources across a range of physical regimes using different modelling approaches. While these studies have provided valuable insights, certain choices—such as two-dimensional setups or the exclusion of magnetic fields—limit the development of key dynamical processes. In particular, the growth of fluid instabilities is known to be suppressed in two-dimensional simulations \citep{2016A&A...596A..12M}.

\begin{table*}
	\centering
	\caption{List of simulations and parameters}
	\label{tab:sims}
	\begin{tabular}{lcccccccr}
		\hline
		Simulation    & $P_{j}$           & $\gamma_{b}$   & $\sigma_{B}$&$B_{0}$    & $T_{s}$    &$n_{p}$        & Restart      & $T_{r}$\\
            label         & $ (\mbox{erg}\cdot \mbox{s}^{-1})$        &                &    &($\mu$G) & ($\mbox{Myr}$)      &               & label$^{a}$        & ($\mbox{Myr}$)\\
		\hline
		A             & $10^{45}$         & $5$           & 0.1&$629$& $0.46$   &44             & ARe          & $2.28$\\
                          &                   &         &       &                  &              &             & ARl               & $7.29$\\
		B             & $10^{45}$         & $5$             & 0.05&$445$& $0.46$   &42            & BRe               & $2.43$\\
                          &                   &        &        &                  &              &             & BRl               & $7.31$\\
		C             & $10^{44}$         & $2$         & 0.05&$142$& $1.37$    &16             & CRe               & $4.25$\\
                          &                   &        &        &                  &              &             & CRl               & $12.8$\\
		\hline
	\end{tabular}
    \begin{tablenotes}
        \item[] $^{a}$The labelling convention is as follows: In ARe, A stands for the parent simulation, R stands for restart, and e stands for early restart. Similarly, 'l' and the end of the simulation label stand for late restart.
        \item[]$P_{j}$: Jet power calculated from simulation.\\
        \item[]$\gamma_{b}$: Bulk Lorentz factor of the jet at injection.\\
        \item[]$\sigma_{B}$: Jet magnetisation parameter, the ratio of jet Poynting flux to enthalpy flux at the jet injection.\\
        \item[]$B_{0}$: Injected magnetic field strength as defined by Eq.~16 of \cite{2020MNRAS.499..681M}.
        \item[]$T_{s}$: The time at which the first jet is stopped.\\
        \item[]$n_{p}$: Number of Lagrangian macro-particles per cell
        \item[]$T_{r}$: The time at which the second jet is launched.\\ 
    \end{tablenotes}
\end{table*}
To address these limitations, we perform three-dimensional simulations of restarted jets with properties representative of typical FR--II sources \citep{1974MNRAS.167P..31F}. Our setup is designed to capture the complexity of restarted jet evolution, with a focus on quantifying how entrainment modifies the remnant cocoon and whether it can decelerate a subsequent jet episode. The three-dimensional setup enables the development of instabilities and associated mixing, while the inclusion of magnetic fields allows us to quantify their influence on these processes. The computational domain extends up to $50\,\mbox{kpc}$ in the vertical direction, allowing us to follow the long-term evolution of the system. At the same time, a uniform spatial resolution of $40\,\mbox{pc}$ ensures that small-scale structures and mixing are well resolved, while avoiding artefacts associated with non-uniform grids.

The jet-cocoon dynamics are treated within a relativistic framework, which is essential for accurately capturing the jet propagation, shock formation, and energy transport. A self-consistent inclusion of magnetic fields enables us to study their effects on stabilising the flow, regulating the growth of instabilities and mixing. We also incorporate an external gravitational field to understand the roles of buoyancy and gravitationally triggered instabilities in shaping the morphology and evolution of the cocoon after the jet turn-off. 

We employ Lagrangian tracer particles to sample the subgrid evolution of the local nonthermal particle distribution, using a physically motivated prescription for shock acceleration, thereby allowing us to study nonthermal particle cooling, acceleration and corresponding emission features. The inclusion and self-consistent evolution of magnetic fields, along with subgrid sampling of nonthermal particle distribution, together allow us to calculate synchrotron spectra. The results on the Lagrangian particles and emission signatures will be presented in a subsequent publication. 
 \begin{figure}
    \centering
    \includegraphics[width=0.85\linewidth]{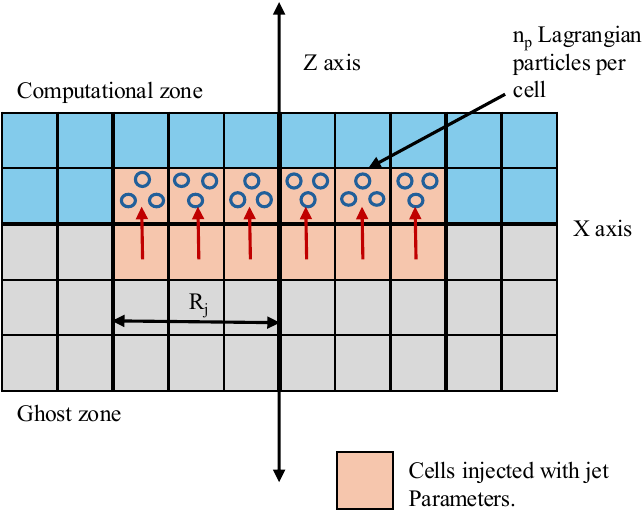}
    \caption{Schematic illustration of the jet injection method. The blue and grey cells represent computational and ghost zones, respectively. The jet nozzle is marked by orange cells.}
    \label{fig:jinj}
\end{figure}
This paper is structured as follows. In Section 2, we present the relevant details of the numerical setup. We describe the initial and boundary conditions, the methods used to restart the jet, and the parameter choices. In Section 3, we present the results, followed by discussion and conclusion in Sections 4 and 5, respectively. 

We use the terms "restarted jet" and "relaunched jet" interchangeably to refer to the second jet following an initial episode of jet activity. We denote the speed of light using $\mbox{c}$. Unless otherwise specified, speeds are measured in units of $\mbox{c}$. Vector quantities are denoted by a boldface letter (e.g. $\boldsymbol{B}$) and their magnitude is denoted by an uppercase letter (e.g. $B$). Average quantities are denoted by angled brackets ($\langle B \rangle$).

\section{Numerical Framework}
We model the restarted jet as a pre-established inflow launched into an ambient medium that is initially in hydrostatic equilibrium within the gravitational potential of the host galaxy and its dark matter halo. We evolve the jet until it reaches $20\; \mbox{kpc}$ and a reflective boundary condition is applied at the jet injection region (jet nozzle), effectively stopping the jet injection. The cocoon created by the initial jet is then allowed to evolve passively. A new jet is then relaunched into the remnant cocoon at a pre-determined time. For each initial jet, two jet relaunches are performed at different stages of cocoon evolution.
The evolution of the system is governed by the equations of relativistic magnetohydrodynamics, which are listed in Appendix~\ref{app:eqn}. These equations with appropriate initial and boundary conditions are solved using the {\small PLUTO} code \citep{2007ApJS..170..228M}.

\subsection{Computational domain and numerical methods}
The simulations are performed over a uniform grid of size $650\times650\times1250$ cells with a resolution of $40\;\mbox{pc}$. The $x$ and $y$ axes extend from $-13$ to $13\;\mbox{kpc}$ while the $z$ axis spans $0$ to $50\;\mbox{kpc}$. Thus, the origin is situated at the centre of the lower boundary ($z=0$). This makes the physical domain cover physical lengthscales of $26\times26\times50\;\mbox{kpc}$ in the x, y, and z directions, respectively. 

To solve the RMHD equations, we utilise the piecewise parabolic scheme \citep{2014JCoPh.270..784M} with a second-order Runge-Kutta method for time integration and the HLLD Riemann solver \citep{2009MNRAS.393.1141M} from the {\small PLUTO} code. The face-centred magnetic field components are evolved using the constrained transport method \citep{1999JCoPh.149..270B,2004JCoPh.195...17L,2005JCoPh.205..509G} which ensures $\nabla.\boldsymbol{B}=0$ up to machine level accuracy. Electromotive force defined on cell edges is evolved using the CT CONTACT scheme\citep{2005JCoPh.205..509G}. 

For numerical stability, strongly shocked regions, as well as the region near the jet injection zone, are solved using a more diffusive HLL Riemann solver and linear reconstruction with MINMOD limiter. To overcome occasional numerical issues, we employed the more diffusive UCT HLL scheme \citep{2003A&A...400..397D, 2004JCoPh.195...17L} to average the electromotive force during the cocoon decay phase in the low-power simulation (Sim~C, see Table~\ref{tab:sims} for simulation details).

\subsection{Initial and Boundary Conditions}
We initialise the simulation domain with an ambient medium that is in hydrostatic equilibrium in the host galaxy halo potential following \cite{2020MNRAS.499..681M}. Except for the $z=0$ surface, where the jet is launched from the central region, all other boundaries follow a restricted outflow boundary condition. On these boundaries, matter is allowed to flow out, but inflows are numerically restricted. This precaution is taken to avoid contamination of the numerical domain due to spurious inflows when the shock created by the jet reaches the domain boundary.

The $z=0$ boundary is set to be reflective except for the jet injection region. As shown in Fig.~\ref{fig:jinj}, the jet is injected by assigning requisite jet injection parameters at ghost cells just below the jet nozzle. Additionally, the first computational cells just above the jet nozzle are also assigned the jet injection parameters when the jet is active. We apply a radial smoothing profile to the jet velocity at the nozzle to avoid sharp discontinuities between the injected jet and the ambient medium:
\begin{equation}  
g(r)=
    \begin{cases}
     \frac{1}{\rm{cosh}\left[\left(\frac{4r}{3r_{j}}\right)^{6}\right]} & r\leq r_{j}\\
     0                                                                     & r > 0.
    \end{cases}
\end{equation}
This ensures that the mass, momentum, and energy fluxes across the jet nozzle at $z=0$ result in a jet of the power $P_{j}$ (listed in Table~\ref{tab:sims}) being launched into the simulation domain. The jet is switched off when it reaches
$20\;\mbox{kpc}$. This is implemented by smoothly transitioning the jet nozzle from outflow (injection) boundary conditions to a reflective boundary over a few computational time steps. Since this transition spans only a few time steps, the total time during which the jet is turned off is dynamically irrelevant. Thus, the jet shutdown is effectively instantaneous in physical terms. After the shutdown, the entire $z=0$ boundary is set to reflective as mentioned before. Once the jet is inactive, the computational cells immediately above the nozzle (see Fig.~\ref{fig:jinj}) are no longer prescribed jet parameters and instead evolve self-consistently according to the equations of RMHD.

\subsection{Summary of Simulations}

We have performed three simulations of the initial jet, referred to as Sim~A, Sim~B, and Sim~C, with varying jet power and magnetic field strength to isolate and study the impact of these parameters on the evolution of restarted jet systems. We adopt jet powers of $10^{44}$ and $10^{45}\,\rm{erg s^{-1}}$, consistent with estimates for FR--II radio galaxies \citep{2010ApJ...720.1066C}. The jet power is controlled by varying the bulk Lorentz factor, $\gamma_{b}$, which takes values of 3 and 5 for the low- and high-power cases, respectively. These values lie within the range inferred from Doppler boosted luminosities of Blazars \citep{2007ApJ...658..232C, 2009AJ....138.1874L}.

In all simulations, the jet radius is fixed at $r_{j} = 200\;\mbox{pc}$. The density contrast between the jet and the ambient density, defined by $\eta = n_{j}/n_{a}$, is set to $10^{-4}$ at the jet launch ($z=0$). This corresponds to a jet number density of $n_{j} = 10^{-5}\;\mbox{cm}^{-3}$. The jet pressure is fixed at $p_{j} = 2.29\times10^{-10}\;\mbox{dyn}\;\mbox{cm}^{-2}$. We perturb the transverse velocity at the jet inlet following \cite{2008A&A...488..795R}. Thus, apart from the bulk Lorentz factor $\gamma_{b}$, all fluid variables are identical for the three initial jet simulations. 

Our prescription for injecting magnetic fields is identical to \cite{2020MNRAS.499..681M}. The injected magnetic field is quantified using Poynting flux to  the jet enthalpy flux ratio, expressed through the magnetisation parameter
\begin{equation}\label{eq:sigB}
\sigma_{B} = \frac{B_{0}^{2} v_{j}}{4\pi F_{j}}.
\end{equation}
Here, $F_{j}$ denotes the jet enthalpy flux, $v_{j}$ denotes jet velocity, while $B_{0}$ represents the magnetic field strength, as defined in Eq.~9 and Eq.~16 of \cite{2020MNRAS.499..681M}, respectively. The choices of $\sigma_B$ and the resulting $B_{0}$ for the three simulations are listed in Table~\ref{tab:sims}. They lead to large-scale magnetic fields in the range of $1-100\,\mu G$ as the jet evolves, which are consistent with FR--II radio sources \citep{2021ApJ...916...95I}. A summary of jet parameters is presented in Table~\ref{tab:sims}.
\begin{figure*}
    \centering
    \includegraphics[width=1\linewidth]{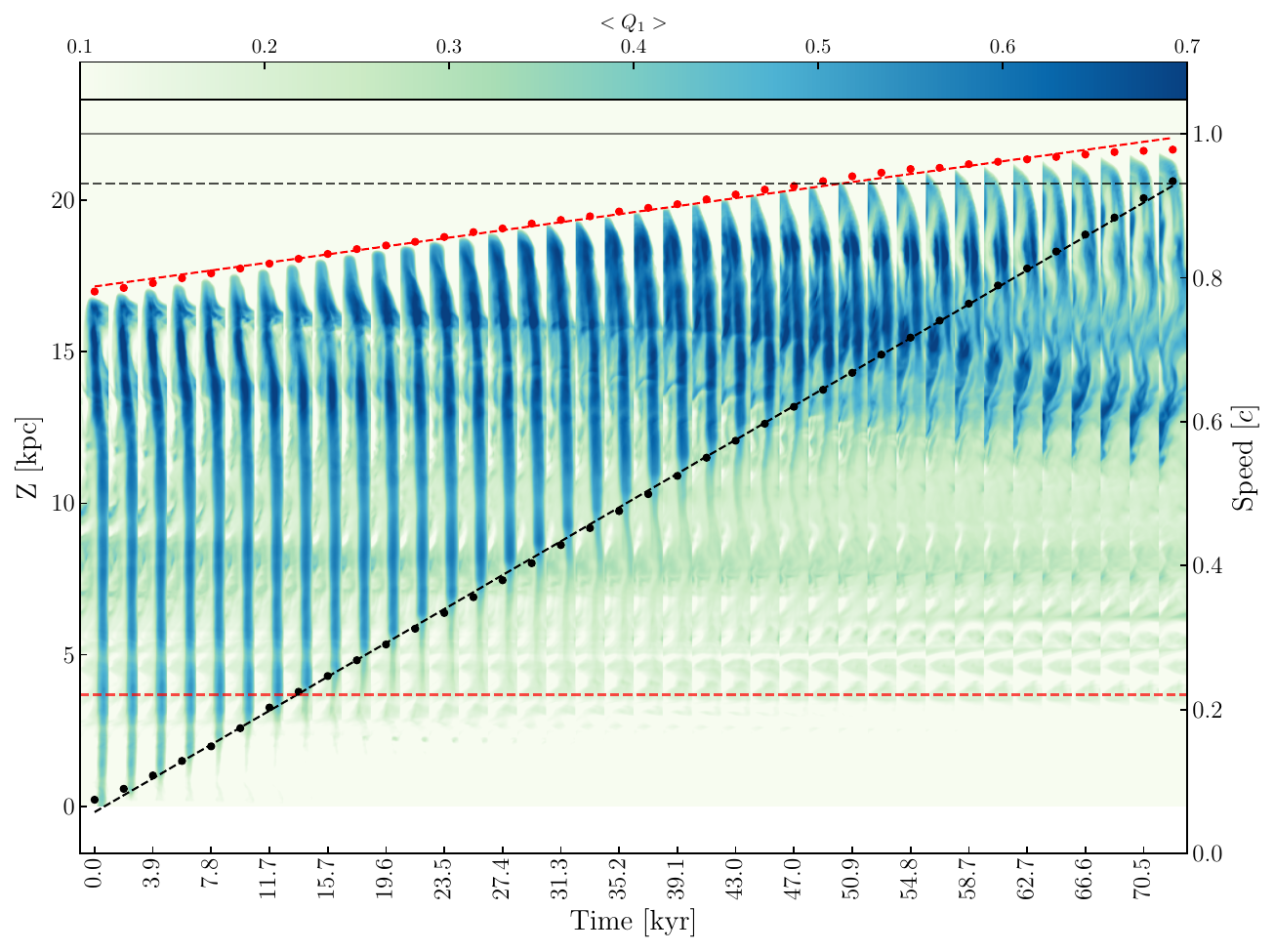}
    \caption{The scatter points represent the location of Jet head (red) and LJM (black) identified using averaged jet tracer $\langle Q_{1}\rangle$, overplotted with a linear fit for Sim~A. The twin axis represents corresponding speeds, shown by dashed horizontal lines with the same colour convention. The background shows slivers of jet tracer ($\langle Q_{1} \rangle$), illustrating the progressive disruption of the jet beam.}
    \label{fig:jet_beam}
\end{figure*}

For each initial jet, we simulate two independent restarted jet episodes. The restart time, $T_{r}$, adopted in each case is listed in Table~\ref{tab:sims}. In the first scenario, the jet is relaunched during the early stages of cocoon evolution (hereafter, the early restart). In the second scenario (the late restart), the jet is relaunched after the cocoon has undergone substantial dynamical and morphological evolution.

Observational estimates of quiescence times in restarted radio galaxies span a wide range, from very short interruptions \citep{2013MNRAS.430.2137K,2012MNRAS.424.1061K} to several tens to hundreds of Myr \citep{2012A&A...545A..91S,2022A&A...661A..92B,2023A&A...677A...4C}. \cite{2019A&A...622A..17S} found that the most massive elliptical galaxies are always active, favouring short interruption timescales. However, such studies favouring shorter quiescence times may be affected by observational bias, as the emission from the older cocoon fades rapidly once the jet switches off \citep{2020MNRAS.496.1706S}. In our simulations, the jet stop and restart times, $T_{s}$ and $T_{r}$, are selected to ensure that the relaunched jet encounters the cocoon at clearly distinct stages of its dynamical evolution, while keeping the simulation costs within available computational resources. For the adopted parameters, this requires quiescence times longer than the active phase of the initial jet.

\begin{figure*}
    \centering
    \includegraphics[width=1\linewidth]{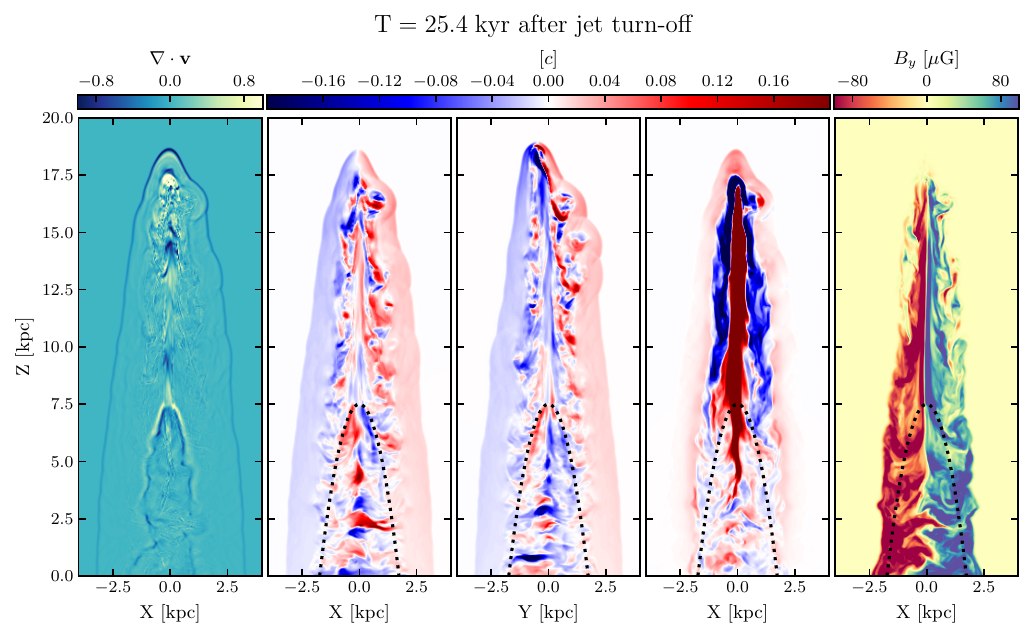}

    \caption{$\nabla \cdot \mathbf{v}$, $v_{x}$, $v_{y}$, $v_{z}$, and $B_{y}$ in the X--Z $(y=0)$ plane shortly after jet turn-off for Sim~A. The jet beam is disrupted up to $\sim 7.5\;\mbox{kpc}$. The outward-propagating compression wave that defines LJM's wake is visible as a down-facing parabolic contour centred at $\sim 7.5\;\mbox{kpc}$ in the first panel ($\nabla \cdot \mathbf{v}$). An approximate outline of this wave is plotted on the subsequent panels using dotted lines. The $v_{x}$ and $v_{y}$ plots show strong lateral flows within the wake of the LJM. The next panel ($v_{z}$) shows an intact jet beam above $7.5\;\mbox{kpc}$ along with typical backflow. The final panel shows a more disturbed magnetic field topology within the jet's wake as compared to the region above the LJM.}
    \label{fig:jet_wake}
\end{figure*}
In both the restart scenarios, we use the same injection parameters as in the initial jet. This choice is motivated by an analytically and observationally supported model \citep{2013MNRAS.436.1595K} that favours comparable jet power across multiple activity episodes.

\section{Results}
In this section, we first describe the dynamical evolution of the remnant cocoon, using Sim~A as the reference case. We then examine key aspects of its evolution, including the growth of the cocoon size, the role of buoyancy, the development of fluid instabilities, mass entrainment, and the evolution of the magnetic field. The other simulations are used to isolate the effects of specific physical parameters: Sim~B is employed to investigate the impact of lower magnetisation (compared to Sim~A), while Sim~C is used to assess the role of lower power (compared to Sim~B). By comparing these models, we identify which evolutionary trends are generic and which depend on the jet properties. In Section~\ref{sec:relaunch}, we study relaunched jets propagating through the remnant cocoons to examine how the cocoon conditions established in the earlier evolution influence the properties of the restarted jet. This is achieved by relaunching the jets at two different times during the evolution of the remnant cocoon. Finally, we discuss the dynamics as the jet erupts out of the cocoon in section~\ref{sec:breakout}.
\subsection{Dynamical phases of remnant cocoon}
\subsubsection{First jet and turn-off}\label{subsec:ljm}
 Relativistic jets propagating into a homogeneous, stratified ambient medium have been studied extensively (see Section 2.1 of \cite{2025Galax..13..102M}). Such jets typically exhibit underdensity and supersonic speeds relative to their surroundings. The jets decelerate upon impact with the ambient medium, creating a forward (bow) shock. The deflected jet material at this shock creates a backflow. The resulting structure consists of dense shocked ambient medium (SAM) enveloping backflow of shocked jet material, separated by a contact discontinuity (CD). Together, these regions form the cocoon. We use the terms SAM and outer cocoon interchangeably for shocked ambient material. We use the terms cavity and inner cocoon interchangeably for shocked jet material \citep[see Fig. 1 from][]{2011ApJ...740..100B}.

The inner cocoon, filled with high-pressure backflow, surrounds the entire length of the jet. It exerts lateral pressure on the jet, confining it. The jet adjusts to this external pressure, producing a series of recollimation shocks along its length, especially within the first few kiloparsecs.

The cocoon expands differently along and across the jet axis because the local pressure and density vary with direction. In environments with decaying ambient pressure and density profiles, the jet ram pressure drives rapid vertical expansion. In contrast, the cocoon expands more slowly in the transverse direction, where thermal pressure dominates. As a result, the cocoon takes on a conical shape as reported in previous studies \citep[e.g.][]{2018MNRAS.480.5286Y,2020MNRAS.499..681M}.

The jet–cocoon structure persists while the jet remains active, but the evolution changes once the jet shuts off. We define the last jet material (LJM) as the material injected immediately before turn-off, occupying only a few computational cells. After shut-off, the LJM continues to move upward, and the jet head remains unaffected by the turn-off until the LJM reaches it.

The positions of the jet head and the LJM, which mark the endpoints of the jet column, are identified using the method described in Appendix~\ref{app:tracer}. Fig.~\ref{fig:jet_beam} illustrates the evolution of the length of the jet column post-turn-off. While the jet head (red dots) travels with a sub-relativistic speed of around $0.2\mbox{c}$, the LJM (black dots) travels almost at the speed of light, similar to the findings of \cite{2014MNRAS.439.3969W}. As a result, the LJM catches up with the jet head. When this happens, the entire jet beam is destroyed. Between the cessation of the jet and the LJM reaching the jet head, the jet head moves up from $15\;\mbox{kpc}$ to $22\;\mbox{kpc}$ (see Fig.~\ref{fig:jet_beam}). The LJM catches up with the jet head in time comparable to the light crossing time of the cocoon height at that moment ( $\frac{22\;\mbox{kpc}}{c}\;=\; 71.75\;\mbox{kyr}$).

As the LJM moves upward, it leaves behind an evacuated region that a steady jet would otherwise refill. Instead, cocoon material flows inward to fill this space, consistent with \cite{2014MNRAS.439.3969W}. This rapid infall drives a compression wave and produces localised sub-relativistic motions, which may enhance turbulent mixing within the cocoon. Fig.~\ref{fig:jet_wake} shows snapshots of relevant quantities when the LJM has reached around $7\;\mbox{kpc}$. The outward-moving compression wave is visible in the $\nabla. \mathbf{v}$ plot inside the bow shock as a roughly parabolic contour with apex at $\sim 7\; \mbox{kpc}$. Alternating regions of sub-relativistic flow directed toward the jet axis are evident in the $v_{x}$ and $v_{y}$ panels. The $B_{y}$ distribution highlights a clear contrast across the LJM: the magnetic field ahead of it remains ordered, while the trailing region appears disordered due to the disruption following jet turn-off.
\begin{figure}
    \centering
    \includegraphics[width=0.9\linewidth]{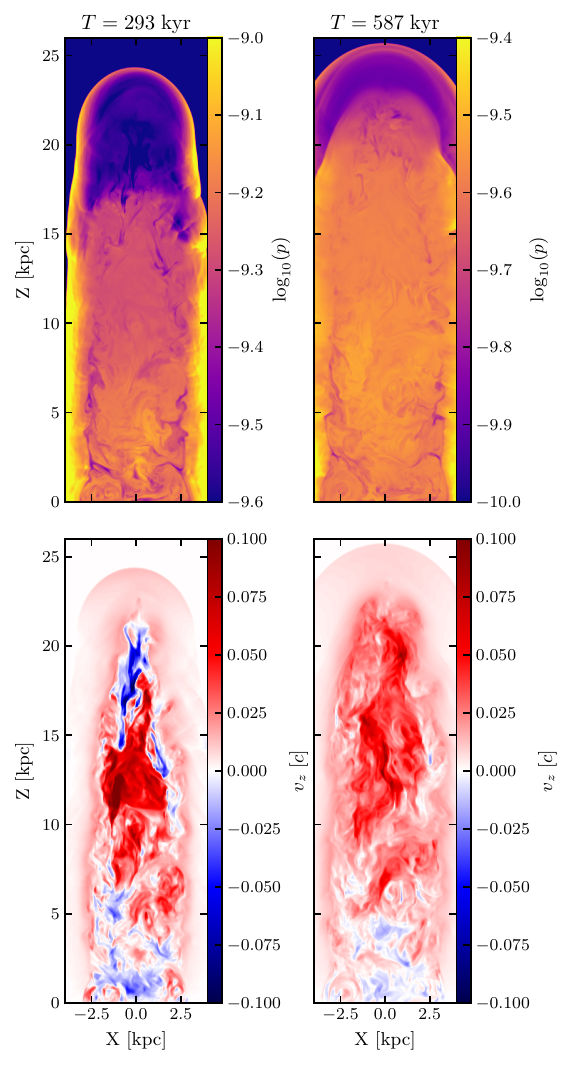}    
    \caption{The top row shows pressure ($p$) in the units of $\mbox{dyn}\;\mbox{cm}^{-2}$, in the X--Z $(y=0)$ plane, at $T=293\;\mbox{kyr}$ and $T=587\;\mbox{kyr}$ after the jet turn-off for Sim~A. The bottom row shows corresponding vertical velocity ($v_{z}$) fields.}
    \label{fig:var_cyl}
\end{figure}
\subsubsection{Cylinder Phase}
At the jet termination shock, the jet ram pressure prevents downstream shocked jet material from moving back along the jet axis. Due to its high pressure, this material expands sideways before being directed downward, forming the backflow \citep[see Fig. 1 from][]{2011ApJ...740..100B}. Once the LJM crosses the jet termination shock, ram pressure support for the downstream shocked jet material abruptly ceases. Hence, the shocked LJM moves directly downward along the jet axis, unlike the sideways then downward motion of a typical backflow. By then, the jet plasma immediately above the LJM has already crossed the shock and contributed to the backflow. As a result, shortly after the LJM crosses the termination shock, downward-moving material populates the top of the cocoon. Such backflow occurs in all simulations, but it eventually quenches; the entire cocoon then rises, as explained below.

After the jet beam is completely disrupted, the forward shock continues to expand due to its prior momentum. This advance remains supersonic (see Fig.~\ref{fig:kh}) throughout the evolution in our simulations, maintaining the pressure jump at the leading edge, as seen in the top row of Fig.~\ref{fig:var_cyl}. However, the CD lags behind the forward shock due to a lack of continuous thrust from the jet. As a result, the gap between them widens, causing the forward shock to detach from CD. Such shock detachment is also reported by other studies \citep{2025PASA...42..152S, 2014MNRAS.445.1462P, 2002MNRAS.336..649K}.

As forward shock detaches and continues to propagate outward, backflows occupy the upper regions of the inner cocoon. These oppositely directed motions produce a low-pressure, evacuated region near the top of the cocoon (Fig.~\ref{fig:var_cyl}, row 1, col 1) which is subsequently filled by high-pressure material rising from the lower cocoon (Fig.~\ref{fig:var_cyl}, row 1, col 2). Once the backflow is quenched, the cocoon is dominated by an upward-moving material as seen from the second row of Fig.~\ref{fig:var_cyl}.

In Appendix~\ref{app:B1}, we examine the effect of stopping the jet at a larger height. In this case, the transient backflow following jet turn-off is more prominent because the taller jet column supplies a larger amount of shocked jet material to the cocoon. Nevertheless, the subsequent evolution remains qualitatively similar: the backflow eventually quenches, and the cocoon becomes dominated by upward-moving material. The main difference is that these evolutionary stages occur over longer timescales.

\begin{figure*}
    \centering
    \includegraphics[width=0.9\linewidth]{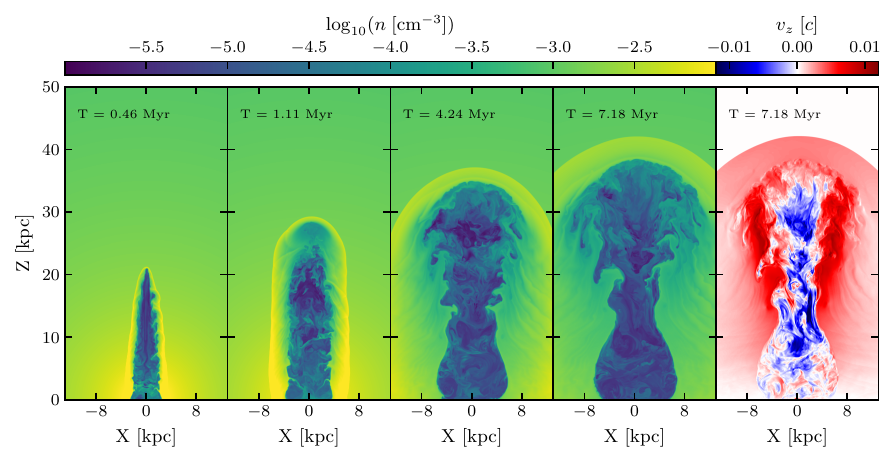}
    \caption{Density slices in X--Z (y=0) plane indicating progressive evolution of of remnant cocoon. The last panel shows the $v_{z}$ distribution corresponding to the previous panel. Labels at the top of each panel represent the time of the slice.}
    \label{fig:phases}
\end{figure*}

\subsubsection{Mushroom Phase}
We define the end of the cylinder phase as the stage when most of the inner cocoon material is moving upward.
After the cylinder phase, the cocoon continues to rise and gradually transitions into a new morphology.  Fig.~\ref{fig:phases} summarises the overall evolution from an initially conical structure (active jet), through the cylindrical phase, to the later mushroom-shaped phase. The last panel of Fig.~\ref{fig:phases} shows that the inner cocoon flow, which was upward-dominated during the end of the cylinder phase, has transitioned to predominantly downward motion.

As the cocoon rises in the stratified atmosphere, its upper regions encounter lower ambient pressure. Hence, the overpressured cocoon undergoes an enhanced lateral expansion near the top, producing a broad upper structure while the lower cocoon remains relatively narrow.

Fluid instabilities play an important role in the evolution of the cocoon during this phase, which we discuss in detail in Sec.~\ref{subsec:inst}.
Velocity shear across the CD drives Kelvin--Helmholtz (KH) instability, producing the large-scale vortices visible in Fig.~\ref{fig:phases}. The 3rd and 4th panels of this figure illustrate the growth of these vortices. They deform the upper region of the cocoon, producing a narrow passage connecting the upper and lower parts. At the top, CD becomes Rayleigh--Taylor (RT) unstable.

The combined effects of differential expansion and KH-driven deformation produce a mushroom-shaped cocoon with a broad head and a narrow stalk. As the cocoon evolves slowly in this phase, this phase dominates the evolution of the remnant cocoon.

\begin{figure}
	\includegraphics[width=0.95\linewidth]{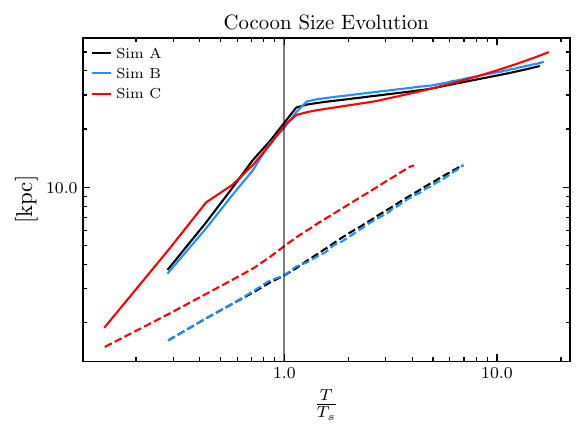}
    \caption{The evolution of the lateral and vertical extent of the cocoon for different simulation cases. The solid and dashed lines represent the height and radius of the cocoon, respectively. Vertical line marks the time of jet turn-off. The radial evolution plots are truncated at a point where the forward shock escapes the simulation domain from the lateral boundaries.}
    \label{fig:sz_evolve}
\end{figure}
 \begin{figure}
     \centering
     \includegraphics[width=0.95\linewidth]{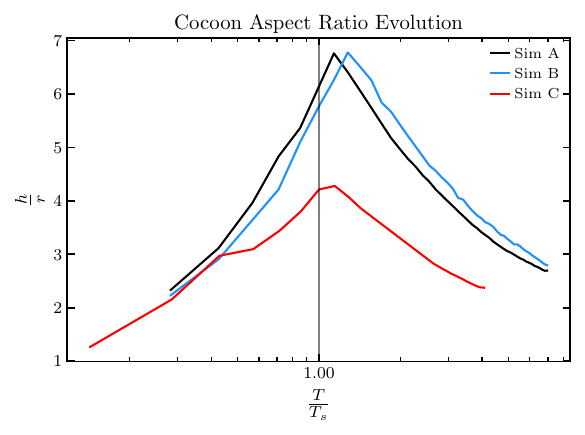}
     \caption{Cocoon aspect ratio evolution for Sim~A, B and C. Vertical line marks the time of jet turn-off.}
     \label{fig:aspect}
 \end{figure}
\subsection{Remnant cocoon geometry}
In this section, we examine the time evolution of the cocoon size and shape. We measure the cocoon height and base radius using a tracer-based criterion described in Appendix~\ref{app:tracer}. Figure~\ref{fig:sz_evolve} shows the evolution of these quantities. The height evolution reflects the different dynamical phases discussed earlier, whereas the radius shows a smoother trend. This difference arises because, at jet turn-off, the driving mechanism for vertical expansion changes from ram pressure to thermal pressure, while the lateral expansion remains thermally driven throughout.

During the active phase, the cocoon expands rapidly along the jet axis compared to the radial direction, resulting in a conical shape. After jet turn-off, the cocoon continues to expand rapidly for a short time, until the LJM catches up with the jet head. The vertical expansion then slows, marking the onset of the mushroom phase. This change in expansion rate produces a clear “knee” in the height evolution. Despite differences in jet power, all simulations show similar trends in both height and radius. This indicates that the ambient medium primarily governs cocoon size evolution.
 
After the jet switches off, thermal pressure drives both vertical and lateral expansion. This common expansion agent suggests that the cocoon should evolve self-similarly, preserving its aspect ratio. We test for self-similarity by tracking the evolution of aspect ratio (height-to-radius ratio) over time (Fig.~\ref{fig:aspect}).

Fig.~\ref{fig:aspect} illustrates the rapid increases of aspect ratio during the active phase as the jet head propagates through an ambient medium with steeply declining density. After jet switch-off, the aspect ratio decreases, with a short delay relative to the shutdown (until LJM catches up with the jet head). In the low-power case (Sim~C), the smaller $\gamma_b$ produces slower vertical growth and limits the maximum aspect ratio to $\sim 4$, compared to $\sim 6$ in Sim~A and B. 

 The conical geometry at switch-off causes the cocoon's top and sides to sample different ambient conditions, leading to anisotropic expansion. Despite the common thermal driving, the aspect ratio does not remain constant post turn-off: The cocoon does not evolve self-similarly. Similar departures from self-similar evolution have been reported in previous studies \citep{2023Galax..11...87T,2018MNRAS.474.3361T,2013MNRAS.430..174H}.

\subsection{Buoyancy}
After the cylindrical phase, the inner cocoon is an underdense bubble enveloped by a denser outer cocoon. Studies have examined the evolution of such remnant cocoons using simulations tailored to specific physical conditions, including highly magnetised configurations with varying field topology \citep{chen23a}, large-scale jets in complex environments \cite{2025PASA...42..152S}, and mixing processes \cite{2002MNRAS.332..271R}. The buoyant rise of such remnant cocoons has been studied numerically by \cite{2001ApJ...554..261C}, where the cocoon is modelled as a uniform, coherent structure subject to global buoyant and drag forces.

\begin{figure*}
    \centering
    \includegraphics[width=0.9\linewidth]{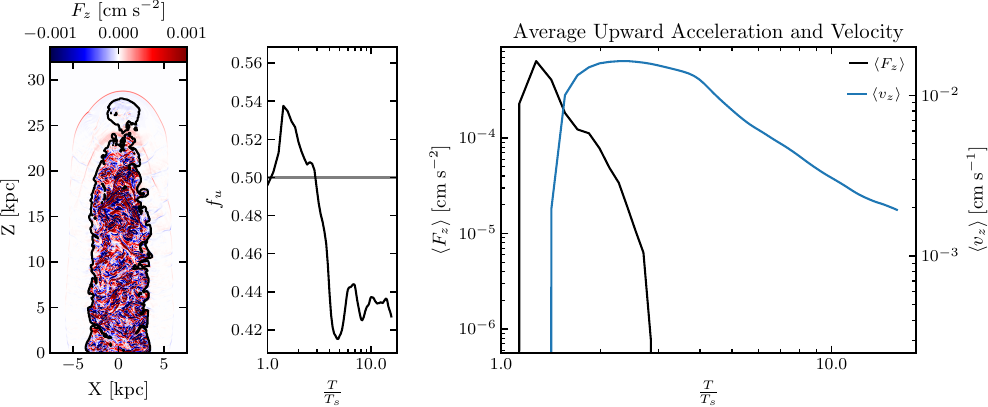}
    \caption{The first panel shows the vertical acceleration, $F_{z}$, for Sim~A during the cylinder phase in the X--Z $(y=0)$. The inner cocoon boundary is indicated by the black contour. The second panel displays the time evolution of the fraction of cells in the inner cocoon with positive $F_{z}$. The third panel shows the time evolution of the inner cocoon-averaged vertical acceleration and vertical velocity. For the middle and right panels, the time axis begins at  $,T/T_{s}=1,$ marking the cessation of jet activity. Because the final panel uses a logarithmic Y-axis scale, negative values are suppressed. }
    \label{fig:buoy}
\end{figure*}

Fluid instabilities entrain SAM into the inner cocoon, rendering it increasingly inhomogeneous. Thus, the classical notion of buoyancy--where an underdense object rises due to hydrostatic pressure gradients in a stratified medium--does not strictly apply in our case. However, individual fluid elements within the inner cocoon experience gravitational force and pressure from surrounding cells. These regions are not embedded in a locally hydrostatic environment, nor are the relevant pressure gradients purely hydrostatic. Nevertheless, these fluid elements experience net accelerations arising from the combined action of pressure gradients and gravity, analogous to buoyant forces. To quantify the role of such \textit{effective buoyancy} in the evolution of the cocoon after jet shut-off, we define the vertical acceleration acting on a fluid cell as

\begin{equation}
F_{z} = -\frac{1}{\rho}\frac{dp}{dz} - g_{z},
\end{equation}
which includes contributions from the pressure gradient and gravity ($g_{z}$). The first panel of Fig.~\ref{fig:buoy} shows that $F_{z}$ is relevant only for the material of the inner cocoon. Depending on the local pressure gradients and density, different regions within the inner cocoon experience either upward or downward acceleration. As a result, although the inner cocoon is underdense, it does not rise coherently as a single structure.

To quantify whether upward or downward acceleration dominates, we define
\begin{equation}
f_{u} = \frac{N_{u}}{N},
\end{equation}
where $N_{u}$ is the number of cells experiencing upward acceleration ($F_{z} > 0$) and $N$ is the total number of cells within the inner cocoon. As shown in the second panel of Fig.~\ref{fig:buoy}, $f_{u}$ exceeds 0.5 only for a short interval after the jet turn-off. During this period, the inner-cocoon-averaged vertical acceleration becomes positive, as seen in the third panel. The average vertical velocity is initially downward at the start of the cylindrical phase. It becomes positive during this interval in response to the net upward acceleration. However, it later decreases as dense, entrained clumps fall under gravity and begin to dominate the flow.

\subsection{Fluid instabilities}\label{subsec:inst}

It has been shown by \cite{1982A&A...113..285N} that a CD with supersonic shear is stable against KH instability. In our simulations, although the jets are highly supersonic, the associated backflows are subsonic to transonic. As the flow decays after jet turn-off, it becomes increasingly subsonic during the cylindrical and mushroom phases. Under these conditions, the CD becomes more susceptible to KH instability, and large-scale modes grow efficiently. Figure~\ref{fig:kh} shows the subsonic inner cocoon flow and the development of KH vortices in Sim~A.

At later times, lateral expansion driven by the pressure contrast between the cocoon and the ambient medium causes the mushroom head to flatten and spread. Its upper surface forms an interface between the dense SAM and the lighter inner cocoon in the presence of gravity. As the flow decelerates, this configuration becomes unstable to Rayleigh–Taylor (RT) modes, and dense material begins to sink into the cocoon.

\begin{figure*}
    \centering
    \includegraphics[width=0.9\linewidth]{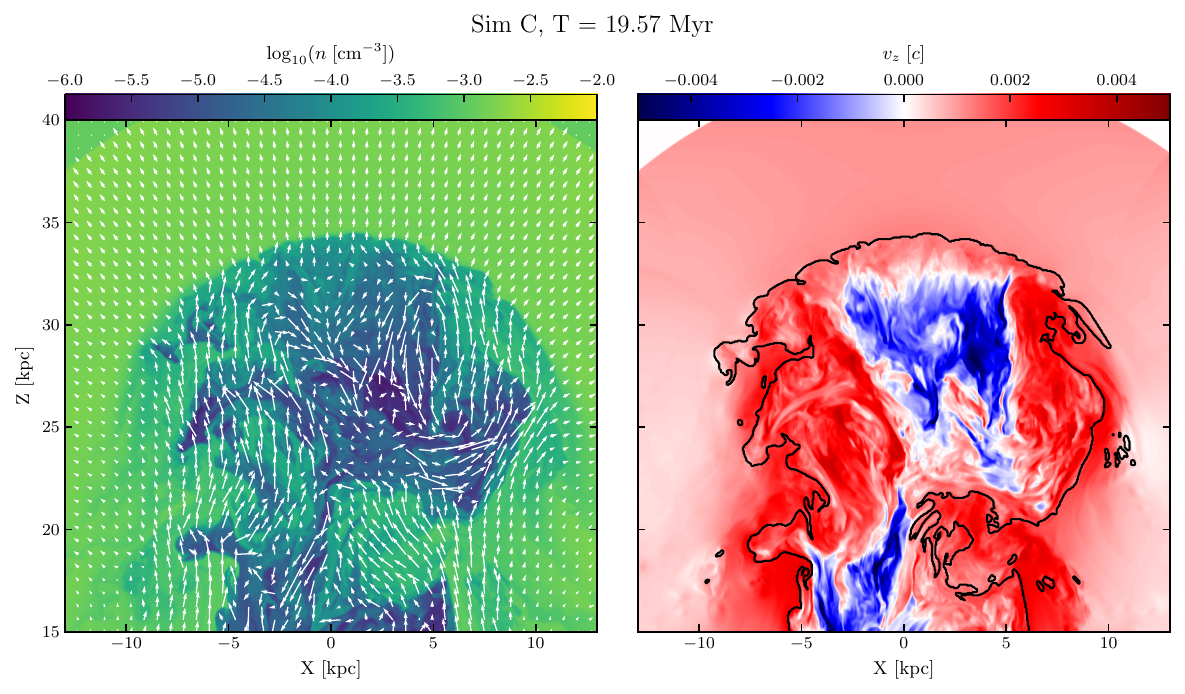}
    \caption{The left panel shows the density field in the X--Z $(y=0)$ for the mushroom phase of Sim~C with the velocity field overlaid. The right panel displays the corresponding vertical component of the velocity, $v_{z}$, showing the RT inflow. The black contour in the right panel delineates the inner cocoon's boundary (CD).}
    \label{fig:vec_v}
\end{figure*}

Figure~\ref{fig:vec_v} shows the density and velocity fields in the upper region during the mushroom phase for Sim~C. The figure highlights both RT-driven inflows at the CD ($v_{z}$ panel) and large-scale KH vortices. The velocity field shows a clear separation: dense material near the boundary moves coherently, while the lighter inner cocoon plasma exhibits disordered, subsonic-to-transonic motion. This turbulence redistributes dense clumps within the cocoon, while coherent inflows form inward-directed vortices that channel material into the cocoon and shape its large-scale structure.

Both KH and RT instabilities drive sustained entrainment of SAM into the inner cocoon, gradually increasing its density. This entrainment contributes to the decay of the buoyant velocity discussed earlier. We examine the entrainment process due to these instabilities in more detail in a subsequent section.

\subsection{The role of magnetic field }\label{sec:mag}
In each simulation, a toroidal magnetic field is injected at the jet nozzle. As the flow evolves, it develops a poloidal component, resulting in a helical field configuration within the inner cocoon that also threads the CD.

\begin{figure}
    \centering
    \includegraphics[width=0.95\linewidth]{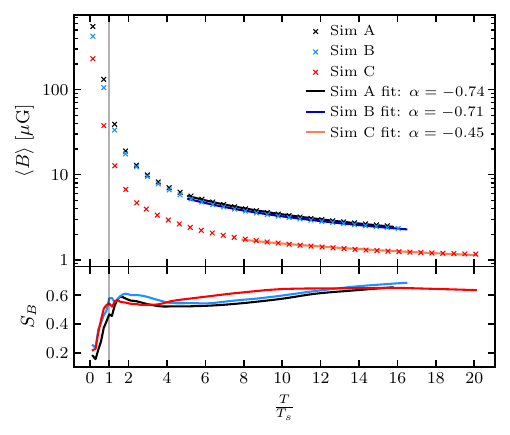}
    \caption{The top panel shows the evolution of the magnitude of the magnetic field averaged over the inner cocoon. Power-law fits to the late-time evolution are plotted over the data. The bottom panel shows the evolution of dispersion of the magnetic field magnitude normalised to the mean ($S_{B}$, see Eq.~\ref{eq:SB}). For both panels, black, blue, and red colours represent Sim~A, Sim~B, and Sim~C, respectively.}
    \label{fig:B}
\end{figure}

Figure~\ref{fig:B} shows the temporal evolution of the volume-averaged magnetic field in the inner cocoon. The method for identification of the inner cocoon for computing the average is described in Appendix~\ref{app:tracer}. During the mushroom phase, the mean field follows an approximate power-law decay,
\begin{equation}
\langle B\rangle \sim \left( \frac{T}{T_{s}}\right)^{\alpha},
\end{equation}
with $\alpha = -0.74$, $-0.71$, and $-0.45$ for Sim~A, B, and C, respectively. Sim~A and B show similar trends, while Sim~C exhibits a shallower decay. Since $B^{2}\propto\sigma_{B} P_{j}$ (Eq.~\ref{eq:sigB}), a factor of two difference in magnetisation corresponds to only a $\sqrt{2}$ change in field strength. As a result, Sim~A and B maintain comparable magnetic field strengths, whereas Sim~C has significantly weaker fields. The slower decay of magnetic fields in Sim~C is consistent with its slower expansion rate.

The inhomogeneity of the magnetic field is quantified by the normalised dispersion,
\begin{equation}\label{eq:SB}
S_{B} = \frac{\sqrt{\langle (B - \langle B \rangle)^2\rangle}}{\langle B \rangle}.
\end{equation}
As seen from the bottom panel of Fig~\ref{fig:B}, during the active phase, $S_{B}$ increases steadily, indicating the growth of turbulent and tangled fields. After jet shut-off, all simulations show only a mild increase, implying that the field remains inhomogeneous rather than relaxing to a uniform state.

We now discuss the effect of the magnetic field on the overall cocoon morphology. Figure~\ref{fig:comp_morph} shows the density distribution when the forward shock reaches $\sim 40\;\mbox{kpc}$. All simulations exhibit a mushroom morphology, but with clear differences. In the low-magnetisation cases (Sim~B and C), large-scale KH vortices form dense walls that separate the head from the stalk. In contrast, these features are less prominent in Sim~A, consistent with magnetic suppression of instabilities. This interpretation agrees with previous studies showing that magnetic fields, particularly with helical structure, inhibit the growth of instabilities \citep{2009ApJ...704.1309D,2021A&A...649A.150B, Elley.M.2026.flickering_jets}.
\subsection{The role of jet power}
\begin{figure}
    \centering
    \includegraphics[width=1\linewidth]{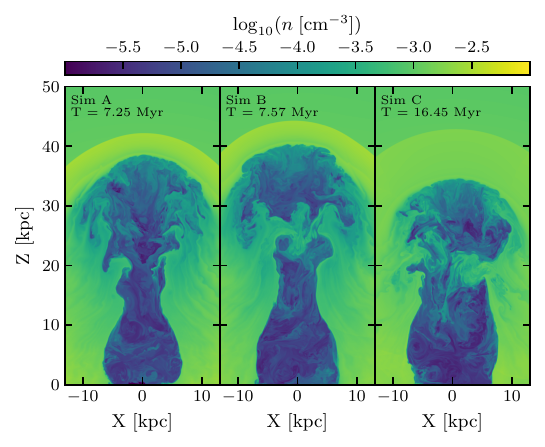}
    \caption{Cocoon morphology at the end of Sim~A, B and C respectively. The first panel shows $v_{z}$ corresponding to the density field in the next panel for Sim~A. Sim~B and Sim~C also show similar spatial variation of $v_{z}$. All panels show slices through the X--Z $(y=0)$ plane.}
    \label{fig:comp_morph}
\end{figure}

In this study, Sim~A and B correspond to jet power of $10^{45}\;\rm{erg\,s^{-1}}$ while Sim~C has a lower power of $10^{44}\;\rm{erg\,s^{-1}}$. These different powers lead to morphological distinctions in the mushroom phase. In Fig.~\ref{fig:comp_morph} we compare the mushroom phases for the three simulations. Except for this dense wall at the neck, Sim~A and B show similar morphologies, whereas Sim~C differs significantly. Due to its lower power, the post–turn-off cocoon in Sim~C has a lower average pressure, resulting in slower expansion. The forward shock takes $\sim 17\;\mbox{Myr}$ to reach $40\;\mbox{kpc}$, compared to $\sim 7\;\mbox{Myr}$ in Sim~A and B.

The lower pressure also limits the vertical advance of the contact discontinuity, producing a thicker outer cocoon as the forward shock detaches and continues to propagate outwards. In Sim~C, the infall of SAM from the top surface (see Fig.~\ref{fig:vec_v}), driven by RT instability, becomes strong enough to nearly halt the upward motion of the CD at $\sim 32\;\mbox{kpc}$. Consequently, the cocoon in Sim~C develops a more disrupted and irregular morphology, with its growth effectively limited by the weight of the overlying SAM. Thus, lower-power jets result in stunted growth of remnant cocoons.

\subsection{Mass entrainment}
Despite the mixing between the inner and outer cocoons, a pronounced density contrast between the two is maintained even during the mushroom phase. We distinguish inner and outer cocoon material using tracer-based criterion described in Appendix~\ref{app:tracer}. The solid lines in Fig.~\ref{fig:entrn} show the time evolution of the mass enclosed within the inner cocoon ($M_{e}$). The dashed lines in this plot indicate the total mass injected by the jet till the time indicated on the $\mbox{X}$ axis. This injected mass, $M_{j}$, is calculated using
\begin{equation}
    M_{j}(T) = \int_0^T \left(\sum_{i}\rho_{i}v_{i}\Delta A\right)\,dT.
\end{equation}
Here, index $i$ runs over the cells of the jet nozzle, yielding mass injection rate with $\rho$, $v$, and $\Delta A$ being the density, velocity and area of the cell, respectively. Although we impose a uniform mass injection, the initial curvature of the dashed lines ($M_{j}$) is attributed to the initial transient phase when the jet is setting up. These curves saturate after the jet is turned off, as expected. 

The total mass within the inner cocoon is much larger than the injected mass since the beginning. A substantial increase in the enclosed mass is observed due to the entrainment of dense outer cocoon into the inner cocoon. Fig.~\ref{fig:entrn} also shows the temporal variation of volume-averaged tracer $\langle Q_{1}\rangle$ within the inner cocoon using dotted lines. The identification of the inner cocoon for the volume average is the same as that for $M_e$ given in Appendix~\ref{app:tracer}. During the active jet phase, the approximately constant value of $\langle Q_{1}\rangle$ indicates a small but steady and fixed contribution of jet material into the enclosed mass. However, the increasing incorporation of entrained SAM leads to a systematic decline in $\langle Q_{1}\rangle$ post-jet turn-off, reflecting the growing dominance of ambient material within the inner cocoon.
  \begin{figure}
     \centering
     \includegraphics[width=1\linewidth]{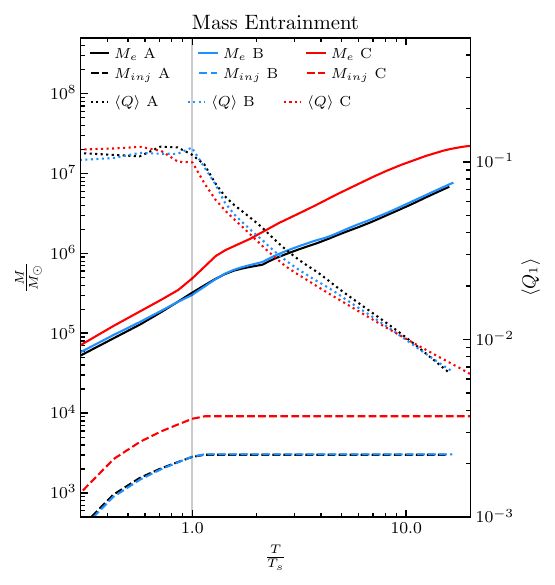}
     \caption{The dashed lines indicate the total mass injected by the jet $M_{j}$ till a given time. The solid lines give the evolution of total mass within the inner cocoon ($M_{e}$) created by the initial jet. The twin axis and dotted lines represents the time variation of the average jet tracer within the inner cocoon. All masses are normalised to the solar mass $M_{\odot}$.}
     \label{fig:entrn}
 \end{figure}

 \begin{figure*}
    \centering
    \includegraphics[width=1\linewidth]{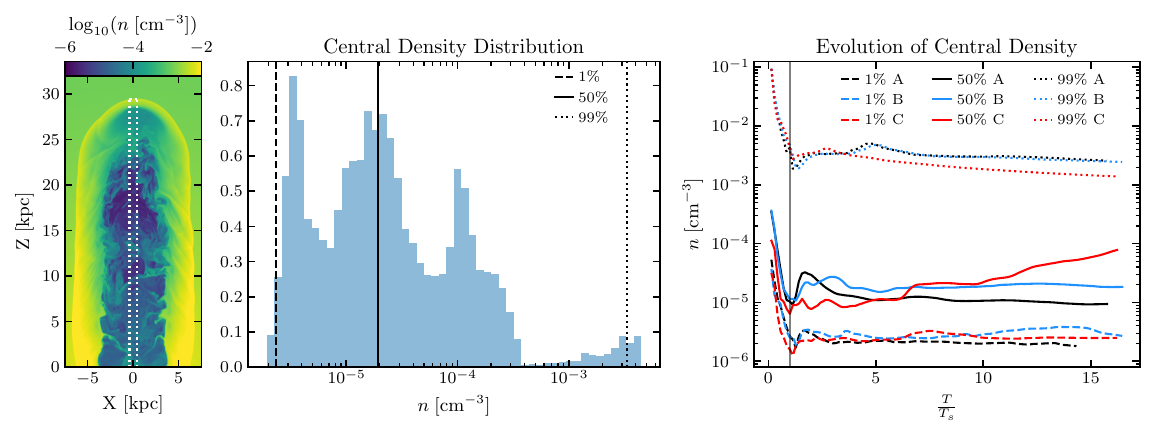}
    \caption{The first panel shows the density field for Sim~A during the cylinder phase in the X--Z $(y=0)$ plane. The second panel displays the density distribution in the central region, marked by the dotted white lines in the previous panel. The vertical lines in the second panel indicate the density that defines the 1st percentile (dashed), median (solid), and 99th percentile (dotted) of the distribution. The time variations of these three markers of the distribution are presented in the last panel, up to the mushroom phase.}
    \label{fig:rho_med}
\end{figure*}

Fig.~\ref{fig:rho_med} captures the time evolution of the probability distribution function (PDF) of density measured along the central column of the inner cocoon. Although the mass entrainment happens mainly at the surface of the inner cocoon, the flattening (or upward trend in the case of Sim~C) of the median density indicates that the entrained material is efficiently distributed throughout the inner cocoon by turbulent flows. Thus, dense structures are present even along the central column of the inner cocoon, providing obstacles for a restarted jet to interact with.

\subsection{Second jet, relaunch}\label{sec:relaunch}

In this section, we study the restarted jet launched into the remnant cocoon of the previous jet. The properties of the ambient medium strongly influence the jet and differ between the initial and restarted jets. A key difference between the initial and relaunched jets is that the initial jet propagates through a much denser ambient medium, whereas the relaunched jet travels through the comparatively rarefied inner cocoon of the earlier jet. Moreover, although the old cocoon as a whole remains over-pressured with respect to the external medium, the pressure at the jet injection site is lower than the central halo pressure encountered by the initial jet. These contrasting density and pressure conditions lead to pronounced differences in the dynamical evolution of initial and restarted jets. 

To study the effect of remnant cocoon evolution, restarted jets in each simulation are launched with identical parameters as the initial jet, but at two distinct epochs: an early and a late restart. In the early case, the jet is injected when the remnant cocoon from the previous epoch is still in the cylinder phase, while in the late case, it is launched into a fully developed mushroom cloud. This ensures that the restarted jets in the early and late cases sample distinct environmental conditions.
\begin{figure*}
\centering 
\includegraphics[width = 0.9\linewidth]{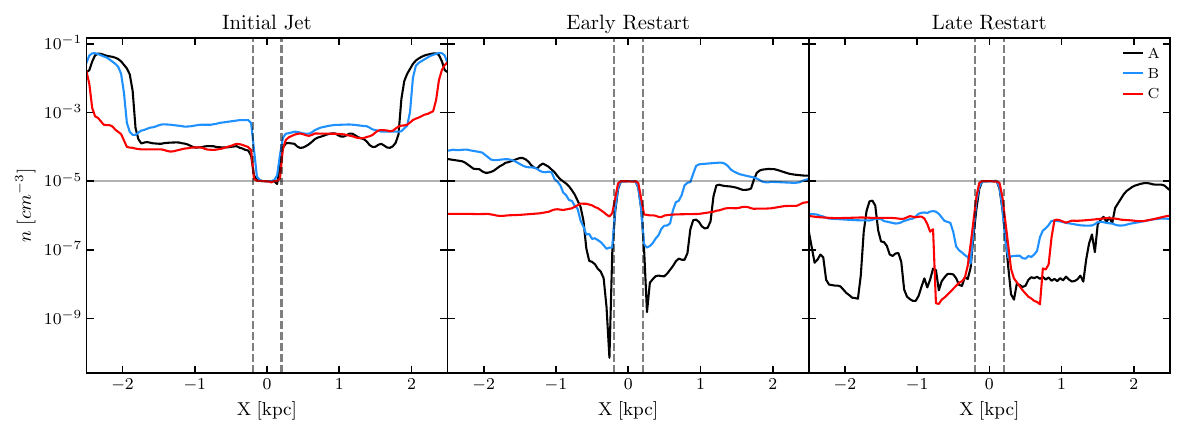} 
\caption{ Density profiles across jet nozzle at arbitrary times along X direction for the initial jet, early restart, and late restart. The vertical dashed lines mark the jet nozzle. The horizontal line represents the injection value of jet density ($n_{j} = 10^{-5}\;\mbox{cm}^{-3}$). } 
\label{fig:overden} 
\end{figure*} 

\begin{figure} 
\centering 
\includegraphics[width = 1\linewidth]{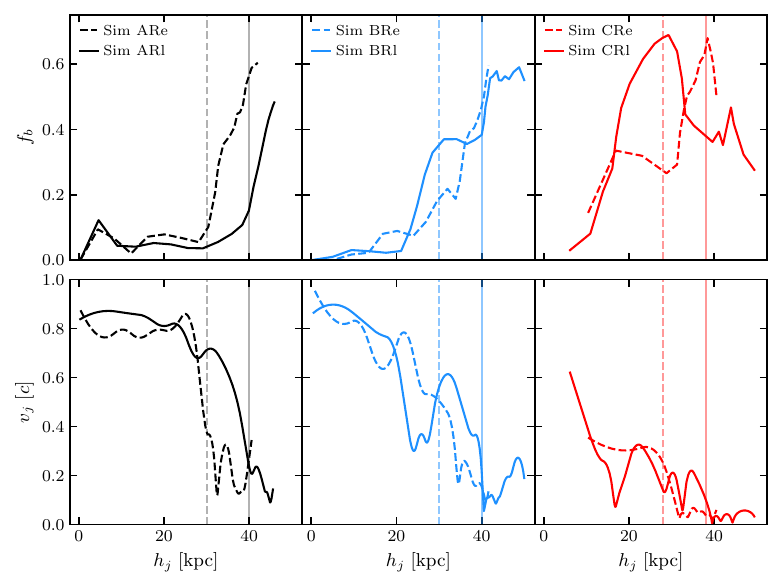} \caption{The top row shows the backflow fraction $f_{b}$ and the bottom row shows the speed of jet head $v_{j}$ as a function of jet length $h_{j}$ for different restarted cases. The vertical lines in each case mark the height of the remnant cocoon.}
\label{fig:j2v}\label{fig:bkf_frac}
\end{figure}
\paragraph*{Backflow, jet advance speed: }The influence of the ambient density contrast at the jet head on the formation of backflow has been studied by \cite{2010MNRAS.405.1303A}. They show that a sharp density jump at the jet head triggers its development, with steeper density gradients producing a more pronounced effect. The density gradient at the jet head is determined by the jet and the ambient density. 

Fig.~\ref{fig:overden} shows density across the jet nozzle along the X direction for all the initial and restarted jets. As seen in this figure, although the initial jets are underdense, the restarted jets are overdense relative to their surroundings in all cases, with the Sim~C restarts showing the least overdensity. Consistent with \cite{1991ApJ...369..308C}, we find that the enhanced over-density allows the jet to propagate ballistically through the remnant cocoon. As a result, the backflow is weak, except when the jet encounters localised regions of higher density.

To quantify the strength of the backflow, we define the backflow fraction as
\begin{equation}\label{eq:fb}
f_{b} = \frac{N_b}{N},
\end{equation}
where $N_b$ is the number of simulation cells within the forward shock of the relaunched jet that have downward velocity, and $N$ is the total number of cells within the same region. The procedure for identifying cells within the forward shock is described in Appendix~\ref{app:tracer}, which also outlines the method used to calculate the restarted jet–head advance speed, $v_{j}$.

The evolution of $f_{b}$ and $v_{j}$ as a function of jet height, $h_{j}$, is shown in Fig.~\ref{fig:j2v}. A correlation between the two can be observed in all the cases. In the restarted cases of Sim~A, the jet propagates ballistically with minimal backflow and an advance speed of $\sim 0.8\;\mbox{c}$ until it breaks out of the cocoon. This ballistic behaviour is a consequence of the highly rarefied inner cocoon and the minimal entrainment observed in Sim~A (see last panel of Fig.~\ref{fig:rho_med}).

Owing to substantial entrainment of dense material into the inner cocoon, the restarted cases of Sim~C exhibit markedly different behaviour from Sim~A. In these simulations, the jet advances more slowly and develops strong backflow as it propagates through the cocoon. This enhanced backflow is most prominent near the peak of $f_{b}$ at $\sim 25\;\mbox{kpc}$ in CRl, corresponding to the jet crossing the dense neck of the mushroom-shaped cocoon.
 
 The restarted cases of Sim~B display behaviour intermediate between Sim~A and Sim~C. The jets initially propagate with high $v_{j}$ and weak backflow, but as they encounter denser structures within the cocoon, $f_{b}$ steadily increases and $v_{j}$ declines. In Sim~BRl, the jet’s passage through the dense neck of the mushroom-shaped cocoon produces a sharp rise in $f_{b}$ and a rapid drop in $v_{j}$ at $\sim 30\;\mbox{kpc}$.

\begin{figure*}
    \centering
    \includegraphics[width = 1\linewidth]{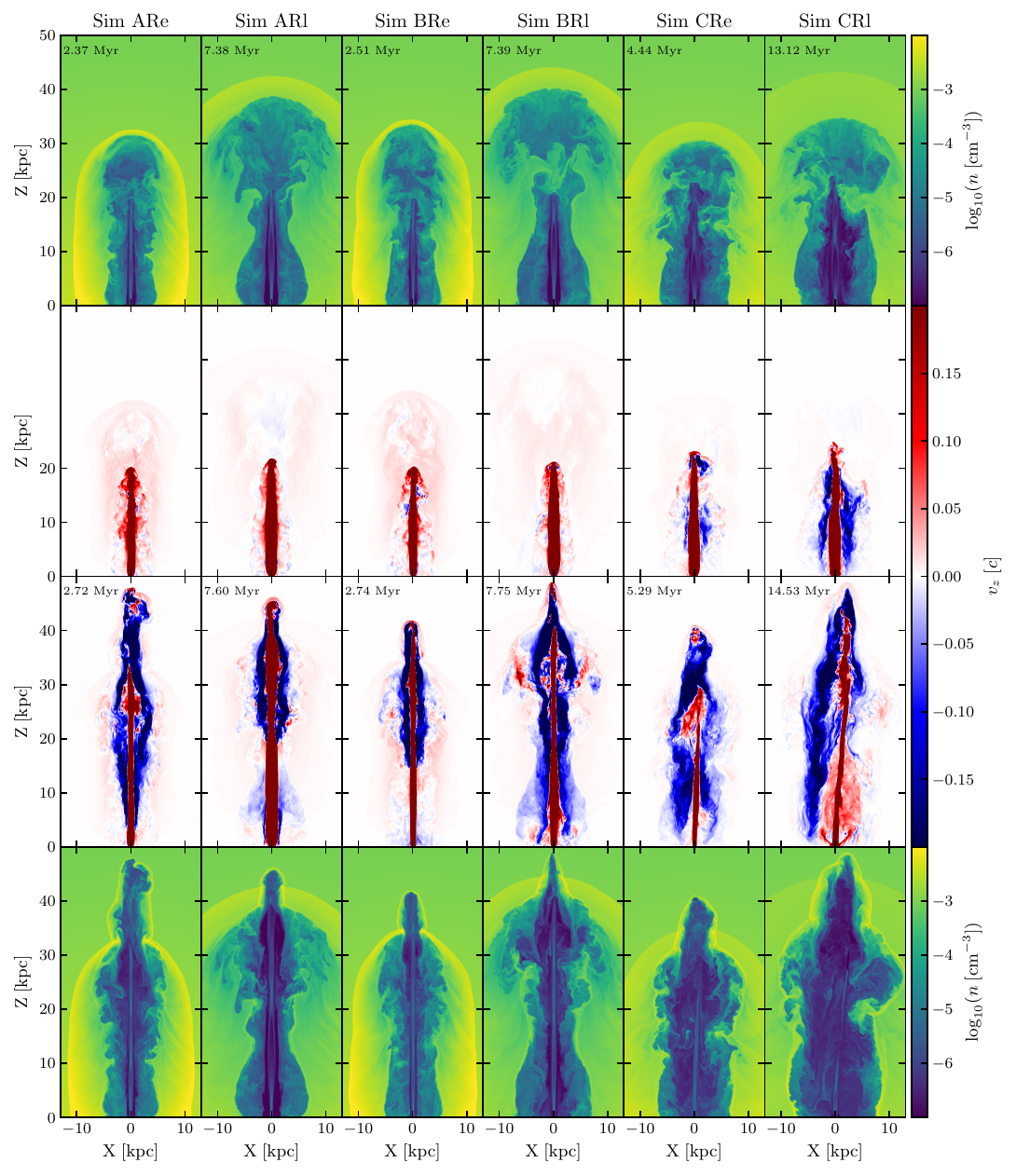}
    \caption{Each column illustrates successive stages in the evolution of different restarted jet cases. The top row shows the density field of the cocoon when the jet has propagated to a distance of approximately $20\ \mbox{kpc}$. The second row displays the corresponding vertical velocity field. The third row shows the velocity field after the restarted jet has erupted from the cocoon. The bottom row presents the corresponding density field at the same stage, after the jet has erupted. Numbers in the top-left corner of the plots indicate the snapshot time. All panels show slices through the X--Z $(y=0)$ plane.}
    \label{fig:summary}\label{fig:late_bkfl}
\end{figure*}

Fig.~\ref{fig:late_bkfl} illustrates the evolution of the vertical velocity $v_z$ and density fields for all restarted simulations. The second row shows that only in Sim~CRe and Sim~CRl does the jet encounter sufficient density contrast while traversing the stalk of the mushroom to generate strong backflows. In all simulations, however, well-developed backflows are established after the jets break out of the cocoon, as seen in the third row of Fig.~\ref{fig:late_bkfl}. 

\begin{figure*}
\includegraphics[width=0.9\linewidth]{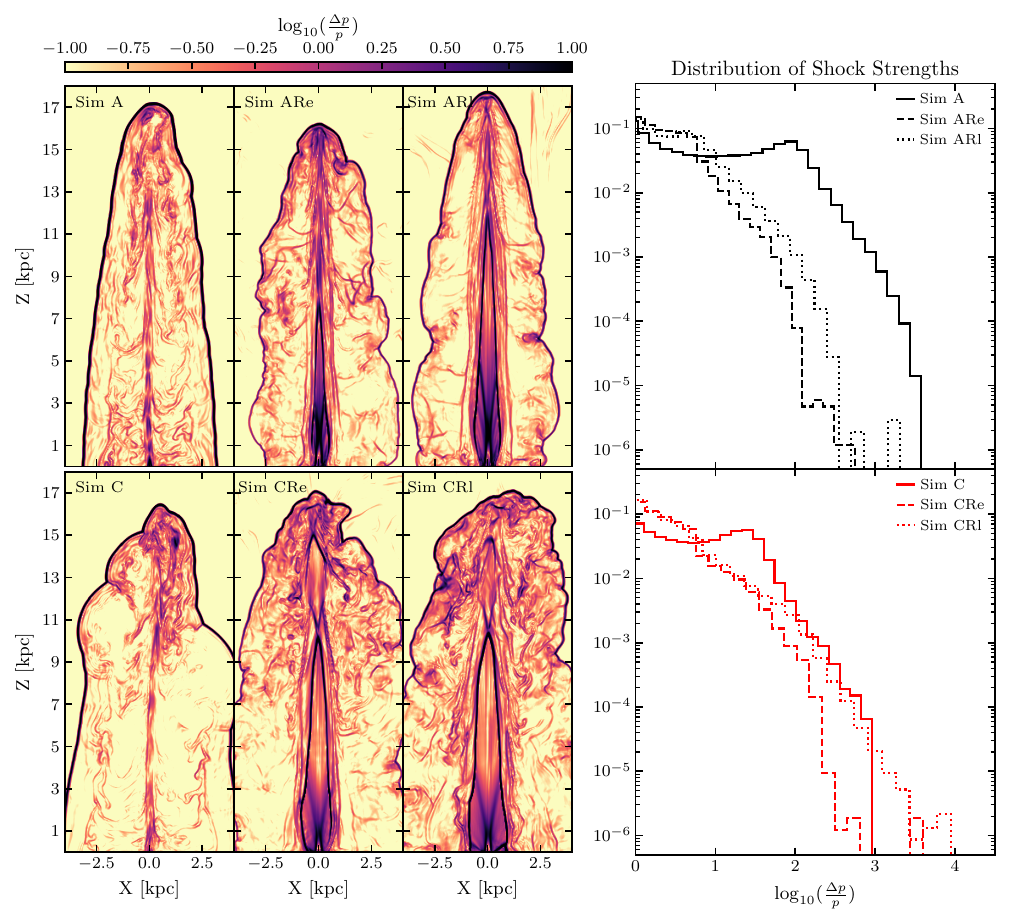}
\caption{Comparison of shock strength and structure for initial and restarted cases in Sim~A and Sim~C. The figure is divided into two major blocks:
\textbf{Left block:} Shock structure quantified using the pressure jump, $\frac{\Delta p}{p}$. The top row displays the $\frac{\Delta p}{p}$ field in the X--Z $(y=0)$ plane when the jet has propagated up to approximately $17\;\mbox{kpc}$ for different cases of Sim~A, while the bottom row shows the same for Sim~C.
\textbf{Right block:} Top panel displays the distribution of $\frac{\Delta p}{p}$ for different cases of Sim~A, and the bottom panel shows the same for various cases of Sim~C.}
\label{fig:delp_p}
\end{figure*}

\paragraph*{Shock structure and strength: }The properties of the ambient medium influence not only the propagation of the restarted jet but also the structure and strength of the shocks it generates. We quantify the local shock strength using the ratio $\Delta P / P$, where $\Delta P$ is defined as the maximum pressure difference between a given cell and any of its neighbouring cells. Fig.~\ref{fig:delp_p} shows the shock structure and strength, represented by $\log_{10}\left(\frac{\Delta P}{P}\right)$, for different jet episodes in Sim~A and Sim~C when the jets have reached similar height. Here we present results only for Sim~A and Sim~C, as Sim~B exhibits trends similar to those of Sim~A. While the shock structures differ markedly across episodes, the cocoons' aspect ratio remains similar within a given simulation set. The side panel of Fig.~\ref{fig:delp_p} illustrates the distribution of the strength of shocks. Examination of the tail of the shock strength distribution shows that, in Sim~A, the strongest shocks are produced during the initial jet phase. Among the restarted cases, ARe and ARl, the latter generates stronger shocks than the early restart due to enhanced entrainment and lower ambient pressure. 

In contrast, the evolution in Sim~C is dominated by substantial entrainment. As a result, both restarted cases exhibit shock strengths that are significantly higher than those of the initial jet, with the late restart (CRl) producing the strongest shocks overall. There is a notable difference in the location of the highest strength shock as well. While Sim~ARe and ARl show the highest strength at the bow shock, in Sim~CRe and CRl, the recollimation and jet-termination shocks are also strong. 

Fig.~\ref{fig:delp_p} also illustrates the shock structures associated with different jet activity episodes. All restarted jets show significantly stronger flaring than the initial jet. This occurs because the relaunched jets are overpressured relative to the remnant cocoon. Although the pressure at the nozzle is the same for both the initial and restarted jets, the surrounding pressure conditions are different. The cocoon as a whole remains overpressured with respect to the ambient medium. However, the initial jet propagates through an atmosphere in hydrostatic equilibrium, with a central pressure much higher than that within the cocoon. As a result, the restarted jets flare more, leading to more pronounced recollimation shocks.

\begin{figure} 
\centering
\includegraphics[width=0.9\linewidth]{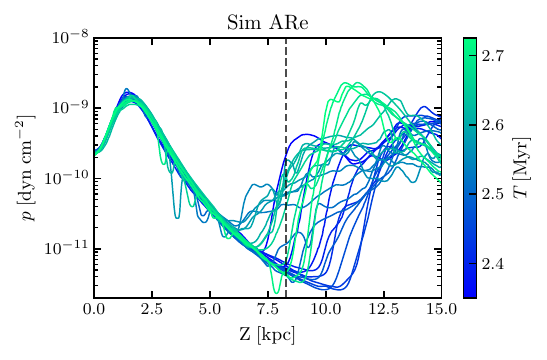} \includegraphics[width=0.9\linewidth]{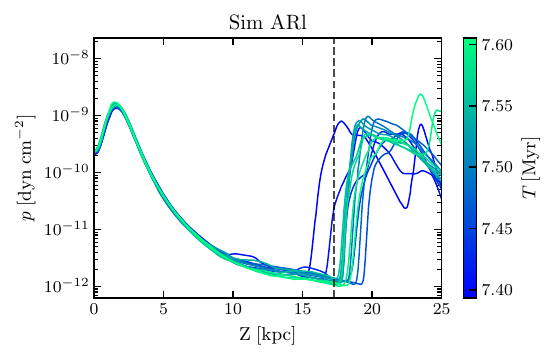} 
\caption{Pressure profile along the jet axis at different times during the relaunch state. The vertical green line represents the height of recollimation shock computed using Eq.~\ref{eq:recol}. The colorbar denotes[] the time corresponding to each pressure profile.} 
\label{fig:recol}
\end{figure}

\paragraph*{Recollimation shocks: }As the cocoon continues to expand after jet shut-off, its pressure gradually declines. Consequently, the location of the first recollimation shock moves to higher heights as the restart time $T_r$ increases. The altitude of the initial recollimation shock, therefore, correlates with the restart delay. The expected height of the recollimation shock can be estimated using Eq.~47 of \cite{1998MNRAS.297.1087K}, given by
\begin{equation}\label{eq:recol}
z_{r} \simeq 6.44\left( \frac{P_{j}}{10^{46}\;\rm{erg\;s^{-1}}}\right)^{\frac{1}{2}}
\left( \frac{p_{c}}{10^{-9}\;\rm{dyne\;cm^{-2}}} \right)^{-\frac{1}{2}}\;\rm{kpc},
\end{equation}
where $P_{j}$ is the jet power and $p_c$ is the pressure of the surrounding medium. We compute the average cocoon pressure in Sim~A just before the launch of Sim~ARe and ARl. Using this average pressure, we estimate recollimation shock heights of approximately $8\;\mbox{kpc}$ and $17\;\mbox{kpc}$ for Sim~ARe and ARl, respectively. Fig.~\ref{fig:recol} shows the pressure profiles along the jet axis at different times during the restarted phase for these two cases. The recollimation shock height is defined as the axial distance between the two dominant maxima in the pressure profile. Although the pressure profiles exhibit temporal fluctuations, once established, the recollimation shock remains close to the height predicted by Eq.~\ref{eq:recol}, as indicated by the vertical green lines in Fig.~\ref{fig:recol}. A comparison between the two cases shows that the early-restarted jet undergoes stronger recollimation, while the later-restarted jet flares more broadly. This difference directly reflects the lower pressure in the inner cocoon at later times, providing weaker confinement for the relaunched jet. A similar broadening of restarted jets in lower-pressure environments has also been reported in hydrodynamic simulations by \cite{2018MNRAS.480.5286Y}.

We now summarise the effect of jet stop time on restart. For the high magnetisation case (Sim~A), even as the cocoon evolves, the mass entrainment is subdominant. Thus, both the restarted jets are ballistic and lack backflows. In cases where entrainment is efficient (Sim~B and C), the later restart shows slower jet propagation speeds and higher backflow. 
Due to decreases in cocoon pressure, the later restarts flare more with higher recollimation shock heights. Later restarts show higher shock strengths than early restarts in all cases.

\subsection{Jet Breakout}\label{sec:breakout}
The summary of all the restarted jet cases is shown in Fig.~\ref{fig:summary}. The first row of this figure shows mid-plane density slices when the restarted jet has reached around $\sim 20\,\rm{kpc}$ and the second row illustrates the corresponding flow structure. It can be seen that during this stage, minimal to no backflow is observed in Sim~A and Sim~B; however, significant backflow is observed in both restart cases of Sim~C. As seen from the 3rd row of this figure, once the jet erupts out of the cocoon, significant backflow is generated. This backflow, which also has much higher pressure than the cocoon, expands as it moves down into the old cocoon.  The bottom row of this figure shows the density slice when the restarted jet has erupted out of the cocoon. During the breakout, as the jets interact with the denser ambient medium, the backflow increases systematically while the jet-head speed $v_{j}$ drops (see Fig.~\ref{fig:bkf_frac}). The steady backflow thus generated can be noted as underdense compared to the cocoon material. 

\begin{figure}
         \centering
         \includegraphics[width=1\linewidth]{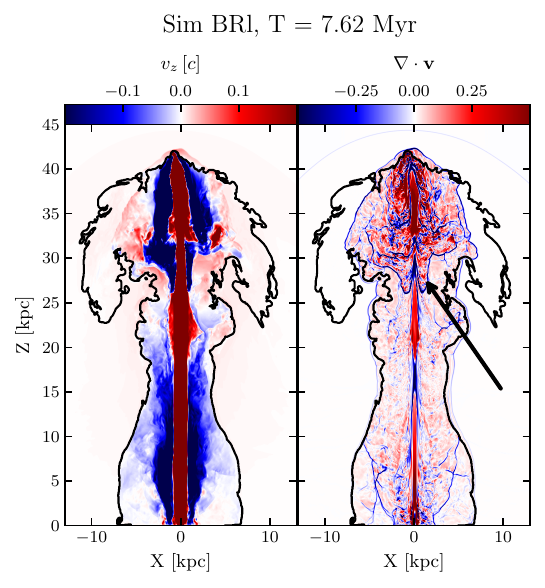}
         \caption{Spatial distribution of vertical velocity $v_{z}$ and $\nabla\cdot \mathbf{v}$ in the X--Z $(y=0)$ plane for Sim~BRl while the jet is erupting out of the cocoon. The back contour marks the boundary of the inner cocoon. The arrow in the second panel indicates the location of the shock created by backflow. }
         \label{fig:BRl_flow}
\end{figure}

We illustrate the complex and rich dynamics of the restarted jets resulting from the remnant cocoon's equally complex evolution using the example of Sim~BRl. Fig.~\ref{fig:BRl_flow} shows the $v_{z}$  and $\nabla \cdot \mathbf{v}$ slice when the jet has started interacting with the outer cocoon in Sim~BRl. Before this, the jet traversed the dense neck region of the mushroom cloud. The temporary backflow produced during that interaction can be seen to be reaching the bottom of the cocoon. Whereas, new backflow is simultaneously excited at the top. This leads to a peculiar flow structure within the cocoon with interrupted backflow. A weak shock is formed at the interaction of the underdense backflow and the dense cocoon medium. This shock is indicated using an arrow in the second panel of Fig.~\ref{fig:BRl_flow} and is observed in other restarted simulations as well (see last row of Fig.~\ref{fig:summary}). The consequences of this shock for the re-acceleration of nonthermal particles and corresponding emission will be discussed in a subsequent publication. 

This example highlights how the internal structure of the remnant cocoon can fundamentally alter the dynamics of restarted jets, producing flow morphologies and shock features that have no analogue in continuously driven systems.

\section{Discussion}\label{sec:disc}
In this paper, we present high-resolution, 3D, RMHD simulations of restarted jets spanning a range of jet powers and magnetisations, with gravity included to model cocoon evolution after jet shut-down more realistically. The simulations reveal rich and complex dynamics of remnant and restarted sources. In this section, we focus on the implications of our results for models of DDRGs, hotspot visibility, and giant radio galaxy formation scenarios. 

\subsection{Is there sufficient entrainment of matter?}\label{subsec:entrn}
\cite{2000MNRAS.315..371S} presented four Mpc-scale radio sources with unusual morphologies, characterised by a pair of outer edge-brightened radio lobes enclosing a smaller, edge-brightened inner pair. Thanks to the advent of new generation surveys, hundreds of such double–double radio galaxies (DDRGs) with closely aligned inner and outer lobes have been identified \citep[see][and references within]{2025A&A...696A..97D}, providing important constraints on the AGN duty cycle and feedback processes.

In a follow-up study, \cite{2000MNRAS.315..381K} (hereafter KSR) developed an analytical model for restarted jets, motivated by the sources discussed in \cite{2000MNRAS.315..371S}, to explain the sizes and emission properties of inner doubles. The model assumes that the hotspot continues to be fed by the jet until the LJM reaches the termination shock, after which it fades instantaneously. The cocoon then expands passively, and a second jet is launched into this relic cocoon to form the inner doubles. Using FRII evolution models from \cite{1997MNRAS.292..723K} and \cite{1997MNRAS.286..215K}, KSR showed that reproducing the observed properties of inner doubles requires substantial additional mass loading of the old cocoon. Fluid instabilities are considered as one of the leading causes of such entrainment, and their role in this process has been widely studied \citep[see][and references within]{2019Galax...7...70P}. Based on characteristic time-scales for the instabilities, KSR favoured the entrainment of warm dense clouds over KH or RT instabilities as the dominant channel for mass loading. Here, informed by our simulations, we revisit the role of fluid instabilities in mass loading of the old cocoon and reassess the conclusions of KSR.

Assuming that the inner cocoon consists solely of jet-injected material, KSR estimates its density as
\begin{equation}\label{eq:ncooc}
n_{c}
= \frac{Q_{o}(t_{o}-t_{d})}{(\gamma_{j}-1) m_{p}c^2 V_{c}},
\end{equation}
where $n_{c}$ is the inner cocoon density, $Q_{o}$ is the jet power, $\gamma_{j}$ is the jet bulk Lorentz factor, $m_{p}$ is the proton mass, and $V_{c}$ is the cocoon volume. Here $t_{o}$ is the total dynamical age of the source, and $t_{d}$ (the dimming time) is the time elapsed since the LJM crossed the jet termination shock. Using this prescription, KSR concluded that the density of the remnant cocoon is too low to explain the observed evolution of inner doubles. Applying Eq.~\ref{eq:ncooc} to our simulations for $t_{o} \sim T_{s}$ (i.e. shortly after jet termination) yields cocoon masses of $2.3\times10^{3} M_\odot$, $2.3\times10^{3} M_\odot$  and $2.7\times10^{3} M_\odot$ for Sim~A, B and C, respectively. These correspond to inner cocoon number densities of $4.0\times10^{-7} \rm{cm^{-3}}$, $4.1\times10^{-7} \rm{cm^{-3}}$ and $2.4\times10^{-7} \rm{cm^{-3}}$. The dashed curves in Fig.~\ref{fig:entrn} show that the mass injected by the jet up to switch-off is consistent with these estimates. However, the solid curves demonstrate that the actual inner cocoon mass—and hence its density—is substantially higher than calculated using Eq.~\ref{eq:ncooc}. Even at later times, as the cocoon continues to expand, the inner cocoon densities seen in our simulations remain higher than would be expected from Eq.~\ref{eq:ncooc}. 

Following \cite{1961hhs..book.....C}, the characteristic growth timescale of KH instability can be estimated as
\begin{equation}\label{eq:tkh}
t_{\rm{KH}}\sim \frac{\sqrt{\chi}\cdot l}{\Delta v},
\end{equation}
where $\chi$ is the density contrast between the SAM and the inner cocoon, $l$ is the perturbation length scale, and $\Delta v$ is the velocity shear across the layer.
\begin{table}
    \centering
    \begin{tabular}{c|c|c|c|c}
    \hline
    & $\chi$ &$l$ &$\Delta v$ &$t_{\rm{KH}}$\\
    \hline
    KSR        &$10^{4}$ &$1\,\rm{kpc}$ &$370\,\rm{km\,s^{-1}}$ &$200\,\rm{Myr}$\\
    This work  & $10$    &$1\,\rm{kpc}$ &$370\,\rm{km\,s^{-1}}$ &$6\,\rm{Myr}$\\
    \hline
    \end{tabular}
    \caption{Comparison of the parameter values used and resulting KH timescale between KSR and this work. Note that all the quantities represent approximate (order of magnitude) estimates.}
    \label{tab:tkh}
\end{table}

In Table~\ref{tab:tkh}, we compare the parameter values and resulting $t_{KH}$ from our simulations with those of KSR. As supersonic shear is stable against instabilities \citep{1982A&A...113..285N}, KSR limits the $\Delta v$ to sound speed of $\sim 370\;\rm{km\,s^{-1}}$. The sound speed in our simulations varies over time within the cocoon. However, the average sound speed in the shear region remains within a range of $\sim 100$ (late time) to $\sim 1000\;\rm{km\,s^{-1}}$ (early time). To ensure a consistent comparison and to isolate the effect of $\chi$, we adopt the same $\Delta v$ as KSR. It should be noted that using the $\Delta v$ from our simulations, especially after the jet turn off, would result in lower $t_{\rm{KH}}$ than reported in Table~\ref{tab:tkh}.

As seen from Table~\ref{tab:tkh}, the smaller  value of $\chi$ calculated from our simulations results in a much lower $t_{\rm{KH}}$ ($\sim 10^{6}\,\rm{yr}$).
 Assuming that the inner cocoon consists solely of jet material leads to a significant underestimate of its density, an overestimate of the density contrast, and hence, an artificially long time-scale for the KH instability ($\sim 10^{8}\,\rm{yr}$; KSR). Our simulations demonstrate that entrainment begins early, even during the active phase of the jet, resulting in lower $\chi$, shorter $t_{\rm{KH}}$, and efficient mixing at the contact discontinuity.

\begin{figure}
    \centering
    \includegraphics[width=1\linewidth]{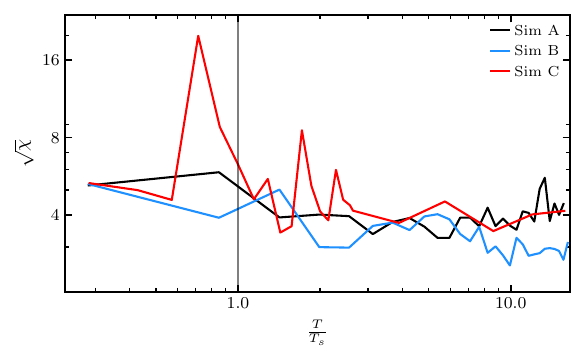}
    \caption{Evolution of $\sqrt{\chi}$ for the three simulations. Vertical line marks the jet turn-off.}
    \label{fig:chi}
\end{figure}
Fig.~\ref{fig:chi} shows the evolution of the maximum $\chi$, which corresponds to the largest KH timescales in each simulation. We compute $\chi$ as the ratio of densities of adjacent cells along the X direction. The figure indicates that $\chi$ ranges from a few tens to a few hundreds, with higher values occurring in Sim~C prior to jet turn-off. After the jet turn-off, $\chi$ remains nearly constant for Sim~A and Sim~C, with a mild downward trend for Sim~B. As entrainment proceeds and the cocoon expands, typical values of $\chi$ decrease, further reducing the KH growth timescale and enabling the growth of larger-scale modes. As seen in Fig.~\ref{fig:comp_morph}, KH modes with spatial scales of several kpc are particularly prominent near the head of the mushroom structure.

Magnetic fields suppress the growth of KH instability. In the presence of magnetic fields, the KH growth rate is given by \cite{2017JPlPh..83f5301F} as:
\begin{equation}\label{eq:taukh}
\tau = \sqrt{\frac{\rho_{1}\rho_{2}(\mathbf{k}\cdot\mathbf{v_{1}} - \mathbf{k}\cdot\mathbf{v_{2}})^{2}}{(\rho_{1}+\rho_{2})^2}
-\frac{(\mathbf{k}\cdot\mathbf{B_{1}})^2 + (\mathbf{k}\cdot\mathbf{B_{2}})^2}{4\pi(\rho_{1} + \rho_{2})}}.
\end{equation}
Here, $\rho$, $\mathbf{v}$, and $\mathbf{B}$ denote density, velocity, and magnetic field, respectively, with subscripts 1 and 2 referring to the two regions across the shear layer. In our case, the wave vector is $\mathbf{k} = (0,0,\frac{1}{l})$, and $\mathbf{B_2} = 0$ since the magnetic field is confined to the inner cocoon. Under these assumptions, Eq.~\ref{eq:taukh} reduces to the KH timescale in the presence of magnetic fields:
\begin{equation}\label{eq:tkhb}
t_{\rm{KHB}} = \frac{1}{\tau} = \frac{ t_{\rm{KH}}}{\sqrt{1 - \frac{B_z^2(1+\chi)}{4\pi \rho_{1}\chi\Delta v^2}}} = f_{m}\cdot t_{\rm{KH}}.
\end{equation}
Thus, magnetic fields modify the purely hydrodynamic KH timescale ($t_{\rm{KH}}$) by a multiplicative factor $f_{m}$. The flow is stabilized when $f_{m}$ becomes imaginary, which occurs if
\begin{equation}\label{eq:fm_stable}
1 - \frac{B_z^2(1+\chi)}{4\pi \rho_{1}\chi\Delta v^2} < 0 \, \Rightarrow\,
B_{z} > \sqrt{\frac{4\pi\rho_{1}\chi\Delta v^{2}}{(1+\chi)}}.
\end{equation}
Therefore, magnetic fields exceeding this threshold suppress KH instability in the inner cocoon, while weaker fields only reduce the growth rate.

We now make an approximate estimate of $f_{m}$ for our simulations using the parameter values listed in the second row of Table~\ref{tab:tkh}. Similar to $\chi$ and $\Delta v$, the magnetic field also varies with location and time. Thus, we utilise the inner cocoon-averaged magnetic field, $\sqrt{\langle B_{z}^2\rangle}$. The average number density near the shear layer varies within the range of $\sim 10^{-2}$ to $\sim 10^{-4}\,\rm{cm^{-3}}$. We choose the logarithmic mean of this range for the density estimate, which results in $\rho_1 \approx 1.673\times10^{-27}\;\rm{g\;cm^{-3}}$.

\begin{figure}
    \centering
    \includegraphics[width=1\linewidth]{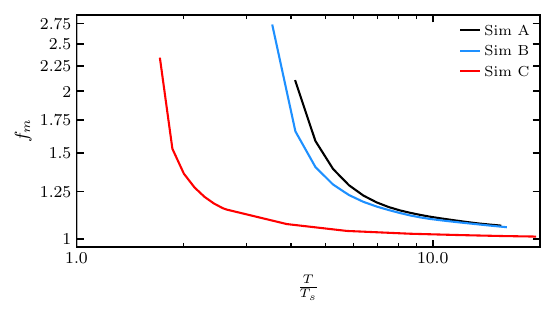}
    \caption{Evolution of $f_{m}$ for the three simulations. The time axis begins at ${T}/{T_{s}}=1$, marking the jet turn off.}
    \label{fig:fm}
\end{figure}

The evolution of $f_{m}$ for the three simulations is shown in Fig.~\ref{fig:fm}. Initially, the magnetic fields are high enough to satisfy the stability criteria from Eq.~\ref{eq:fm_stable}: $f_{m}$ is purely imaginary. Thus, in each case, the curves begin at the time when $f_{m}$ becomes real. This occurs earliest in Sim~C. Initially, $f_{m}\approx2$ in all three cases before gradually decreasing to $\sim 1$ at later times. Sim~A and Sim~B exhibit similar evolution, with the lower magnetisation case (Sim~B) showing slightly lower values of $f_{m}$. Sim~C shows the lowest values of $f_{m}$ at any given time. Lower values of $f_{m}$ indicate weaker suppression of the KH instability in lower magnetisation cases.

Although these approximate estimates illustrate the qualitative behaviour of $f_{m}$, it must be noted that the cocoon is heterogeneous; the parameters that $f_{m}$ depends on vary across the cocoon. Thus, localised regions in Sim~A might have lower $f_{m}$ than some regions in Sim~C. Nevertheless, the average behaviour of $f_{m}$ indicates stronger suppression of KH instability in cases with higher magnetisation.

In conclusion, instabilities at the contact discontinuity can entrain material into the inner cocoon over a few $\rm{Myr}$ timescale. Strong magnetic fields can reduce this entrainment.

\subsection{ When will both the hotspots remain visible?}
Hotspots are compact, bright regions observed at the termination points of powerful radio jets in FR II radio galaxies. They form where the relativistic jet impacts the ambient medium, producing a strong terminal shock that decelerates the flow and efficiently accelerates particles, thereby enhancing synchrotron emission. In a few, typically large restarted sources, hotspots from both episodes are visible \citep[e.g.][]{2024A&A...691A..76W, 2021A&A...649L...6M, 2011MNRAS.414.1397J, 2000MNRAS.315..395S}. This implies that, while the outer hotspot remains actively fed, the inner jet is sufficiently decelerated to form hotspots. Here, we discuss the conditions under which such a scenario is possible.

The disappearance of a hotspot after jet switch-off is not instantaneous. Once the jet is turned off at time $T_{s}$, the hotspot continues to receive jet plasma until the LJM reaches the jet termination point. As discussed in Section~\ref{subsec:ljm}, the LJM propagates at approximately the jet bulk speed ($\sim\mbox{c}$), consistent with findings of \cite{2014MNRAS.439.3969W}, implying that the hotspot remains  active for a time
\begin{equation}\label{eq:hs}
    t_{hs} \simeq \frac{h_{j}(T_{s})}{c},
\end{equation}
where $h_{j}(T_{s})$ is the jet length at the time of switch-off. As noted above, a hotspot fades almost immediately once it is no longer supplied with jet material. The hotspot might remain visible for a short time while it cools due to synchrotron radiation, but the corresponding timescale ($10^4$ yr; KSR) is negligible compared to the dynamical time scales considered here. Thus, Eq.~\ref{eq:hs} provides an upper limit on the post–switch-off hotspot lifetime.

For the hotspot from the first activity episode to be observable simultaneously with a newly formed hotspot from the restarted jet, the inner cocoon must be replenished with sufficient density before the old hotspot fades. This will allow the new jet to produce shocks strong enough to allow efficient particle acceleration. This implies that the entrainment timescale should be shorter than the outer hotspot survival time. The above criterion then leads to
\begin{equation}
    t_{hs} \geq t_{\rm{KH}}, \;\; \mbox{ or: } \;\; h_{j}(T_{s}) \geq c\cdot t_{\rm{KH}},
\end{equation}
suggesting that for a given efficiency of entrainment, longer jet heights at the turn-off would be preferred for the visibility of both hotspots. Using the nominal value of $t_{\rm{KH}}$ from our simulations (see Table~\ref{tab:tkh}) suggests that the first jet must reach Mpc-scale lengths before switch-off for hotspots from both episodes to be visible. 

As shown in Section~\ref{subsec:ljm}, the jet head advances at a fraction of $c$, whereas the LJM propagates at nearly the speed of light. Consequently, the time required for the LJM to reach the jet head ($t_{hs}$) is a fraction of the time required for the initial jet to reach its final size. It follows that the interruption between the two episodes must be a small fraction of the initial jet's active phase, consistent with spectral ageing constraints \citep[e.g.,][]{2012MNRAS.424.1061K}.

These estimates are subject to significant uncertainties. The KH timescale depends sensitively on the density contrast, which evolves as mixing proceeds, generally leading to shorter growth times. In addition, the characteristic velocities and length scales entering the above estimate are uncertain, and our simulations extend only to $\sim 50$ kpc. Therefore, the inferred $\sim$Mpc scale should be regarded as an order-of-magnitude estimate rather than a strict requirement.

More efficient entrainment than captured in our simulations would increase the cocoon density more rapidly, allowing restarted jets to decelerate and form hotspots at smaller source sizes. In such cases, simultaneous visibility of both hotspots may occur in compact restarted sources. However, due to the compact size, the two hotspots might merge on a shorter timescale \citep{2026MNRAS.546ag131E}, making it difficult to observe such sources.

\subsection{How do Giant Radio Galaxies (GRGs) achieve large sizes?}

Galaxies with projected linear sizes in radio emission exceeding $0.7\;\mbox{Mpc}$ are classified as giant radio galaxies (GRGs). These Mpc-scale sources are extreme compared to typical radio galaxies, raising the question of how they sustain accretion and maintain jet collimation and stability over such large distances. The existence of extremely extended systems, including sources approaching several Mpc in size \citep{2024Natur.633..537O}, highlights the need to understand how jets remain stable and collimated over such scales. One possibility is that GRGs preferentially grow in rarefied environments that reduce jet disruption \citep{2015MNRAS.449..955M,2024A&A...686A..21S}, although other studies find no strong environmental dependence on GRG size \citep{2021MNRAS.502.5104L}. Moreover, GRGs larger than $3\;\mbox{Mpc}$ exhibit properties similar to smaller GRGs, suggesting a common evolutionary mechanism rather than a distinct population effect \citep{2025A&A...699A.257A}. Although our simulations probe much shorter length scales, Appendix~\ref{app:B1} and especially Fig.~\ref{fig:B1_sz_evolve} demonstrate that the qualitative evolution of jets switched off at larger heights is similar to that of a jet stopped at lower heights. Thus, the overall dynamical behaviour inferred from our models can be extended to larger scales. 

 As described in Sec.~\ref{subsec:entrn}, magnetic fields can suppress the onset of KH instability. As a result, the entrainment of matter into the inner cocoon will be limited, causing higher values of $\chi$ and longer instability timescales. In high-power, strongly magnetised cases—where the jet is terminated at significantly larger distances, the cylindrical phase may last much longer (see Appendix~\ref{app:B1}). In such cases, the cocoon can remain comparatively underdense for much longer times, with dense material confined to localised structures. This evolutionary pathway is qualitatively similar to Sim~A in our simulations, which develops a relatively rarefied cocoon due to suppression of entrainment by higher magnetic fields.

As shown in Fig.~\ref{fig:j2v}, the relaunched jet in Sim~ARe propagates in an almost ballistic manner while traversing the evolved cocoon. It maintains a high advance speed of $\sim 0.8\;\mbox{c}$. Owing to this ballistic propagation, it crosses the remnant cocoon much faster than the initial jet crossed the ambient medium. As a result, it can reach the height attained by the first jet in a time significantly shorter than the original jet's on-time, quickly “picking up” where the previous episode ended.

This behaviour is qualitatively consistent with the properties of GRGs, e.g. J1420–0545, one of the largest known radio galaxies, spanning several megaparsecs and exhibiting an unusually slim, highly symmetric double-lobed morphology. Detailed modelling indicates that its jets evolve in an extremely low-density environment, enabling efficient, nearly ballistic propagation over large distances \citep{2017ApJ...850....7J}. The high axial ratio and strong collinearity of the radio structure resemble the inner doubles observed in classical double–double radio galaxies, suggesting that the currently visible lobes may correspond to a restarted phase of activity. Independent studies further argue that the exceptional size of J1420–0545 is more plausibly achieved through multiple episodes of jet activity rather than a single, uninterrupted accretion phase \citep{2011ApJ...740...58M}.

Early works hypothesised multiple epochs of activity for GRGs \citep[e.g.,][]{1996MNRAS.279..257S}. Observations suggest that recurrent nuclear activity is important in explaining the large sizes of GRGs. Hard X-ray–selected GRGs often show young or restarted radio cores, indicating re-invigorated accretion \citep{2019ApJ...875...88B}, while soft gamma-ray–selected samples further support restarted jet scenarios \citep{2021MNRAS.500.3111B}. In a population study by \cite{2020MNRAS.494..902B}, 13 of 15 hard X-ray-emitting galaxies exhibit features suggesting multiple jet activity. In another study, GRGs frequently show evidence for restarted jet activity \citep{,2025A&A...696A..97D}. 

Strong magnetic fields may play a dual role in these systems: they can enhance jet collimation and suppress fluid instabilities, keeping the cocoon sufficiently rarefied to permit ballistic propagation, while still maintaining enough residual density to generate strong shocks. These shocks can efficiently accelerate particles, making the restarted jets and their associated structures observable at radio wavelengths.

Mixing between the inner and outer cocoon after turn-off may be insufficient to refill the inner cocoon in high-power, strongly magnetised jets. If GRGs are powered by such jets, incomplete cocoon refilling naturally favours ballistic, stable restarted jets and enables rapid re-extension of the source. In this context, at least a subset of GRGs likely achieve their extreme linear sizes through multiple episodes of jet activity rather than a single, continuous phase. Nevertheless, an important caveat is that very large-scale jets are susceptible to current-driven kink instabilities \citep{2016MNRAS.461L..46T,2021MNRAS.506.1862A,2024JHEAp..44..146U}, an effect that lies beyond the scope of the present study.

\section{Conclusions}
In this paper, we present detailed high-resolution 3D RMHD simulations of jets restarting within the cocoon of a previous episode of activity. We explore how jet power and magnetisation influence the dynamics and evolution of the cocoon during the quiescent phase and examine how these parameters regulate entrainment. In addition, we investigate the impact of the quiescence time, power, and magnetisation on the propagation and interaction of the subsequent jet. The primary conclusions from the fluid dynamics part of the results are:

\begin{enumerate}
    \item The post-switch-off evolution of the system is highly complex and progresses into a mushroom cloud through a cylindrical phase. Buoyant rise plays an important role in this transition.
    
    \item The lateral expansion of the cocoon is largely unaffected by jet switch-off. Following termination, the cocoon aspect ratio decreases. The overall evolution of cocoon height and width is similar across simulations, indicating that the ambient density and pressure profiles primarily control it.
    
    \item As the cocoon evolves, fluid instabilities at the contact discontinuity play a central role in its dynamics. KH instabilities drive substantial entrainment of dense ambient material into the inner cocoon, while at later times RT instabilities further enhance this mixing. The efficiency of entrainment depends strongly on magnetisation: in low-magnetisation cases (Sim~B and Sim~C), instabilities grow more vigorously, leading to higher cocoon densities and more disturbed morphologies, whereas stronger magnetisation suppresses instability growth and limits mixing (Sim~A).

    \item The dense material entrained at the CD is efficiently distributed within the inner cocoon by transonic and subsonic flows. This entrainment of material can explain the deceleration of relaunched jets.
    
    \item Restarted jets propagate approximately ballistically while traversing the stalk of the remnant cocoon, producing minimal backflow. This ballistic motion favours the claim that GRGs achieve large sizes through multiple episodes. A momentary drop in jet propagation speed is observed when the restarted jet interacts with dense structures, creating associated backflow. In the low power and magnetisation case, the remnant cocoon has sufficient density to decelerate the restarted jet, resulting in a typical jet-backflow structure.
    
    \item Restarted jets exhibit enhanced flaring, with the degree of flaring increasing with the quiescent interval. Internal shocks are generally weak, although later restarts produce comparatively stronger shocks (Sec.~\ref{sec:relaunch}). When the restarted jet breaks out of the remnant cocoon, a classical backflow is established. This backflow expands into the under-pressured cocoon, generating a shock.
    
\end{enumerate}

\section*{Acknowledgements}
We thank the anonymous referee for helpful suggestions and comments, which significantly improved the clarity of presentation and improved the overall quality of the paper. We acknowledge support by CINECA through the Italian SuperComputing Resource Allocation (ISCRA) for the availability of high--performance computing resources. The authors acknowledge support from the Indo--Italian project IN22MO08 (INT/ITALY/P-37/2022 (ER) (G)). The authors wish to acknowledge the Inter--University Center for Astronomy and Astrophysics (IUCAA) for the availability of high-performance computing resources and support through the Pegasus Cluster. PR thanks the Sakal India Foundation's Research Scholarship for financial support for travel related to this project. MB acknowledges financial support from Next Generation EU funds within the National Recovery and Resilience Plan (PNRR), Mission 4 – Education and Research, Component 2 – From Research to Business (M4C2), Investment Line 3.1 – Strengthening and creation of Research Infrastructures, Project IR0000034 – “STILES – Strengthening the Italian Leadership in ELT and SKA”, from INAF under the Large GO 2024 funding scheme (project "MeerKAT and Euclid Team up: Exploring the galaxy-halo connection at cosmic noon”) and the Mini Grant 2023 funding scheme (project ‘Low radio frequencies as a probe of AGN jet feedback at low and high redshift’).

\section*{Data Availability}
The simulation data from this research will be shared on reasonable request to the corresponding author.


\bibliographystyle{mnras}
\bibliography{MNRAS/paper1} 




\appendix
\section{Relevant Equations}\label{app:eqn}
In this section, we list out some important equations used in this study.

    \subsection{Fluid equations}
    The evolution of an ideal, relativistic, magnetised plasma is governed by the equations of relativistic magnetohydrodynamics (RMHD), written here in conservative form:
\begin{equation}
    \frac{\partial}{\partial t}
    \begin{pmatrix}
        D     \\
        \boldsymbol{m} \\
        E_{t} \\
        \boldsymbol{B}
    \end{pmatrix}
    + \nabla \cdot
    \begin{pmatrix}
        D\boldsymbol{v}   \\
        w_{t}\gamma^{2}\boldsymbol{v}\boldsymbol{v}
        - \boldsymbol{b}\boldsymbol{b}
        + \boldsymbol{I}p_{t} \\
        \boldsymbol{m}     \\
        \boldsymbol{v}\boldsymbol{B}-\boldsymbol{B}\boldsymbol{v}
    \end{pmatrix}
    =
    \begin{pmatrix}
        0 \\
        \boldsymbol{f}_{g} \\
        \boldsymbol{v}\cdot\boldsymbol{f}_{g} \\
        0
    \end{pmatrix}.
\end{equation}

These equations express conservation of rest mass, momentum, and total energy, along with the induction equation for the magnetic field, in the presence of an external gravitational force $ \boldsymbol{f}_{g} $. The variables appearing in the above equations are defined as follows.  
Here $ \rho $ is the rest-frame mass density, $ \boldsymbol{v} $ is the three-velocity of the fluid measured in the simulation grid frame (SGF), and
\begin{equation}
    \gamma = \frac{1}{\sqrt{1 - \boldsymbol{v}^2/c^2}}
\end{equation}
is the bulk Lorentz factor. The thermal pressure is denoted by $ p $, and $ \boldsymbol{B} $ is the magnetic field in the SGF.
We define
\begin{equation}
    b^{0} = \gamma\,\boldsymbol{v}\cdot\boldsymbol{B},\;     \boldsymbol{b} = \frac{\boldsymbol{B}}{\gamma}
    + \gamma(\boldsymbol{v}\cdot\boldsymbol{B})\boldsymbol{v},
\end{equation}
\begin{equation}
    w_{t} = \rho h + \frac{\boldsymbol{B}^{2}}{\gamma^{2}} + (\boldsymbol{v}\cdot\boldsymbol{B})^{2},
\end{equation}
and the total pressure
\begin{equation}
    p_{t} = p + \frac{1}{2}
    \left( \frac{\boldsymbol{B}^{2}}{\gamma^{2}} + (\boldsymbol{v}\cdot\boldsymbol{B})^{2} \right).
\end{equation}
The conserved variables are
\begin{equation}
    D = \gamma\rho,\;\boldsymbol{m} = w_{t}\gamma^{2}\boldsymbol{v} - b^{0}\boldsymbol{b},\;\mbox{and}\;E_{t} = w_{t}\gamma^{2} - (b^{0})^{2} - p_{t},
\end{equation}
corresponding to the relativistic rest-mass density, momentum density, and total energy density, respectively. The magnetic field satisfies the solenoidal constraint
\begin{equation}
    \nabla \cdot \boldsymbol{B} = 0,
\end{equation}
which is maintained numerically to machine precision.

 \subsection{Equation of state}
The above system is closed using the Taub--Mathews equation of state:
\begin{equation}
    \left( h - \frac{p}{\rho} \right)
    \left( h - 4\frac{p}{\rho} \right) = 1 ,
\end{equation}
where relativistic specific enthalpy, $h$, is given by
\begin{equation}\label{eq:rhoh}
    h = \frac{5p + \sqrt{9p^2 + 4\rho^2 c^4}}{2\rho}.
\end{equation}
This equation of state provides a smooth transition between the relativistic and non-relativistic regimes.

\subsection{Sound speed, mach number}\label{app:mach}
To study the nature of fluid flow in different regimes, we define relativistic Mach number as \citep{2008A&A...488..795R},
\begin{equation}
    \mathcal{M} = \frac{\gamma\cdot v}{\gamma_{s}\cdot c_{s}},\;\gamma_{s} = \frac{1}{\sqrt{1 - \left( \frac{c_{s}}{c}\right)^{2}}}.
\end{equation}
Here, $c_{s}$ is relativistic sound speed, which for the Taub-Matthews equation of state is given by \citep{2005ApJS..160..199M},
\begin{equation}
    c_{s}^{2} = \left( \frac{p}{3\rho h}\right) \left(\frac{5\rho h - 8p}{\rho h - p} \right),
\end{equation}
where $\rho h$ is calculated using Eq.~\ref{eq:rhoh}
\begin{figure} \centering 
\includegraphics[width=1\linewidth]{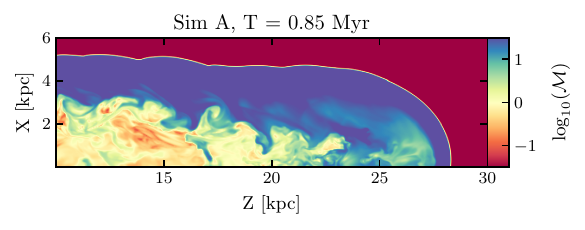} 
\caption{Mach number ($\mathcal{M}$) in the X--Z $(y=0)$ plane showing KH vortices being developed at the contact discontinuity during cylinder of Sim~A as a representative case. Sim~B and C also show a similar spatial distribution of $\mathcal{M}$. Refer to Section~\ref{app:mach} for the definition of $\mathcal{M}$.} 
\label{fig:kh}
\end{figure}

\section{Tracer Variables}\label{app:tracer}

We employ passive tracer variables, $Q_{1}$ and $Q_{2}$, to identify material injected by the initial and the restarted jet, respectively. Each tracer takes values in the range $0 \leq Q_k \leq 1$ and quantifies the fractional contribution of jet material to the local mass density. The ambient medium is initialized with $Q_{1}=Q_{2}=0$ throughout the domain. During the active phase of the initial jet, $Q_{1}$ is set to unity within the jet nozzle, while $Q_{2}$ remains zero. Conversely, during the active phase of the restarted jet, $Q_{2}$ is set to unity in the nozzle and $Q_{1}$ is set to zero.

The tracer variables are evolved as passive scalars using the advection equation
\begin{equation}
    \frac{\partial (\rho Q_k)}{\partial t} + \nabla \cdot (\rho Q_k \boldsymbol{v}) = 0,
\end{equation}
where $k = 1,2$.
In the following subsections, we detail the use of these tracers in the analysis underlying the various results presented in this paper: 
\subsection{Identification of the jet beam}
To identify the jet beam immediately after jet turn-off, we use the tracer $Q_{1}$. We average $Q_{1}$ over a central $24 \times 24$ column in the $X$–$Y$ plane to obtain the axial profile $\langle Q_{1} \rangle$ along the jet direction. The extent of the jet beam is then defined by the outermost points along the $Z$-axis where $\langle Q_{1} \rangle > 0.95$. These extreme points define the jet head and the LJM.

As the LJM propagates through the cocoon, enhanced mixing in its wake makes visual identification difficult. However, the tracer-based criterion provides a robust and consistent method for tracking the LJM throughout its evolution. The motion of LJM and jet head immediately after the jet turn-off is illustrated in Fig.~\ref{fig:jet_beam}

\subsection{Identification of the inner cocoon and cocoon extent}

We identify the inner cocoon boundary, corresponding to the contact discontinuity after jet turn-off, using a combined tracer and density-based criterion. The inner cocoon is defined as the region satisfying $Q_{1} > 0$ and $n < 10^{-3}$, where $n$ is the number density. The tracer condition isolates material with a non-zero contribution from the jet, thereby identifying the cocoon. The density threshold distinguishes the low-density inner cocoon interior from the surrounding dense shocked ambient medium.

\begin{figure}
    \centering
    \includegraphics[width=0.9\linewidth]{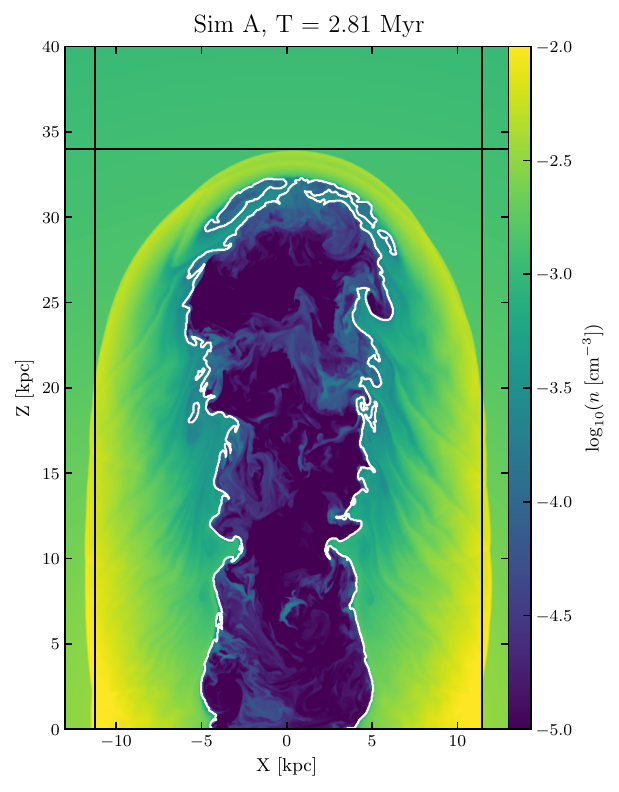}
    \caption{X--Z slice of the density field for Sim~A during the cylindrical phase. White contours mark the inner cocoon boundary, while the vertical and horizontal black lines indicate the axial and lateral extents of the cocoon, respectively. All of these features are identified using a tracer-based criterion.}
    \label{fig:mask}
\end{figure}

The spatial extent of the cocoon inferred solely from the tracer threshold ($Q_{1} > 0$) is illustrated in Fig.~\ref{fig:mask}. The time evolution of the cocoon size, computed from this tracer-defined extent, is shown in Fig.~\ref{fig:sz_evolve}. The inner cocoon boundary obtained using a combined tracer and density criterion is overplotted in Fig.~\ref{fig:mask}. The evolution of the total mass enclosed within this surface, which quantifies the entrainment of dense SAM into the inner cocoon, is presented in Fig.~\ref{fig:entrn}.

\subsection{Speed of the relaunched jet}

To track the propagation of the restarted jet, we use the tracer $Q_{2}$, which uniquely labels material injected during the second jet episode. At each timestep, we identify the uppermost grid cell satisfying $Q_{2} > 0$. The height of this cell defines the instantaneous position of the jet head, $h_{j}$. Differentiating this position with respect to time yields the vertical advance speed of the relaunched jet, $v_{j}$. Evolution of $V_{j}$ as the relaunched jet 

\subsection{Backflow fraction}

The backflow fraction, defined in Eq.~\ref{eq:fb}, is computed using the tracer $Q_{2}$. Let $N$ denote the total number of grid cells satisfying $Q_{2} > 0$, corresponding to the volume occupied by the restarted jet material. The number of back-flowing cells, $N_{b}$, is defined as those cells that additionally satisfy $v_{z} < 0$. The ratio $N_{b}/N$ provides a quantitative measure of the strength of backflow in the restarted jet.

\section{Effect of Jet Stop Height}\label{app:B1}

\begin{figure}
    \includegraphics[width=1\linewidth]{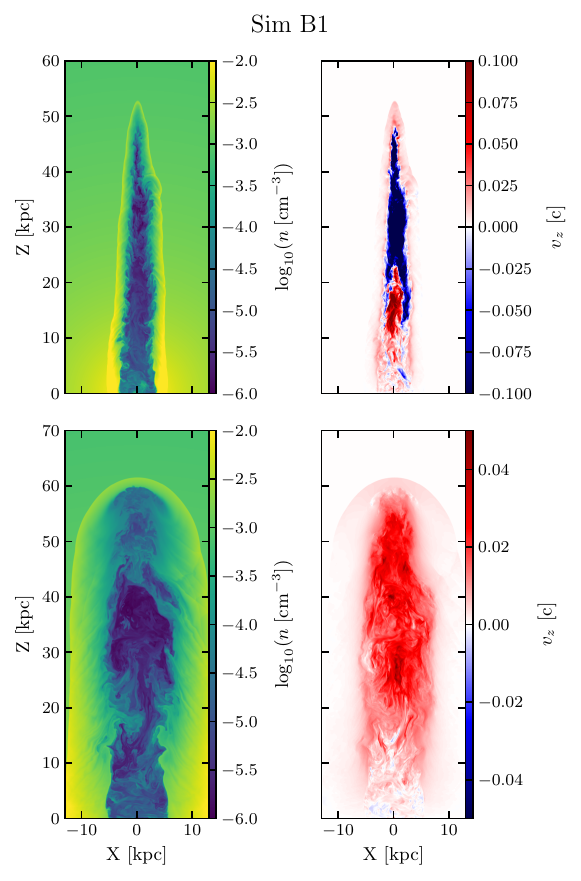}
    \caption{The top row displays the density and velocity structure at the beginning of the cylinder phase for Sim~B1. The bottom row shows the state of these fields at the end of the cylinder phase. The top row and bottom row correspond to $T=0.85\;\mbox{Myr}$ and $T=3.16\;\mbox{Myr}$, respectively. All panels show the fields in the X--Z plane. }
     \label{fig:70kpc}
\end{figure}

In this section, we present results from an additional large-domain simulation aimed at testing whether the qualitative evolution observed in our fiducial models persists for larger jets. In particular, this run examines the sensitivity of cocoon evolution to the jet termination height. It assesses whether restarted jets at greater extents exhibit behaviour similar to that seen in our smaller-scale simulations. This simulation, hereafter Sim~B1, is identical to Sim~B in all respects except that the initial jet is switched off at a height of $30\,\mbox{kpc}$ instead of $20\,\mbox{kpc}$.  Fully capturing the evolution to the mushroom phase in this case would require a substantially larger computational domain. Limited by the computational resources, we evolve Sim~B1 only to the cylindrical phase, as shown in Fig.~\ref{fig:70kpc}.

\begin{figure}
    \centering
    \includegraphics[width=0.9\linewidth]{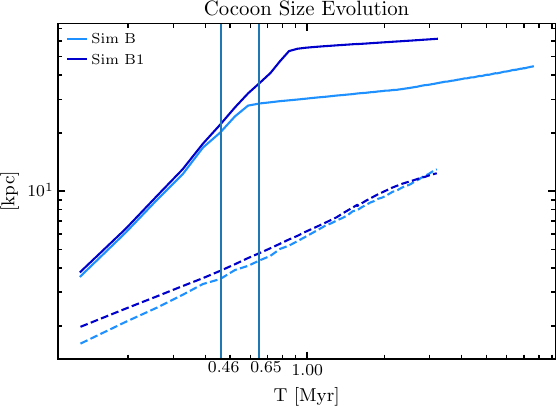}
    \caption{Evolution of the cocoon size in Sim~B and Sim~B1. Solid curves denote the vertical extent (height) of the cocoon, while dashed curves represent its lateral extent. The darker blue curves correspond to Sim~B and the lighter blue curves to Sim~B1. The vertical lines at $0.46\;\mbox{Myr}$ and $0.65\;\mbox{Myr}$ indicate the times at which the jet is switched off in Sim~B and Sim~B1, respectively.}
    \label{fig:B1_sz_evolve}
\end{figure}

The post–turn-off evolution of the cocoon is highly sensitive to the height at which the jet is terminated. This sensitivity arises from the steep decline of ambient density with height: when the jet propagates into a more rarefied medium, its residual momentum decays more slowly. This effect is illustrated in Fig.~\ref{fig:B1_sz_evolve}, which compares the temporal evolution of vertical and lateral extent of the cocoon for Sim~B and Sim~B1. The cylindrical phase persists for a significantly longer duration in Sim~B1. By $T \simeq 4\;\mbox{Myr}$ after jet turn-off, Sim~B already transitions to a mushroom morphology, whereas Sim~B1 still retains a predominantly cylindrical structure. In contrast, the lateral extent of the cocoon, shown by the dashed curves, is nearly identical in both simulations and shows little sensitivity to the jet turn-off.

Fig.~\ref{fig:70kpc} illustrates the beginning and end of the cylindrical phase for Sim~B1. The top row shows the early cylindrical phase, characterised by downward-moving material filling the upper part of the inner cocoon. This flow pattern is more pronounced in this case due to the presence of a longer column of jet that feeds the backflow. The bottom row shows the end of the cylindrical phase, marked by a coherent upward motion of material throughout the cocoon and inner cocoon. The corresponding density fields demonstrate the progressive growth of dense filaments entrained into the inner cocoon during this phase.

To conclude, the larger restarted jets show similar behaviour to the jets presented in this study, albeit at different time-scales.


\bsp	
\label{lastpage}
\end{document}